\documentclass[twocolumn]{article}

\pdfoutput=1

\usepackage[left=0.71in,top=0.95in,right=0.71in,bottom=0.75in]{geometry}
\usepackage{caption}
\usepackage{epsfig}
\usepackage{subfigure}
\usepackage{alphalph}
\usepackage{amsmath}
\usepackage{float}
\usepackage{setspace}
\usepackage{authblk}

\newcommand{\etal}{et al.\@}
\newcommand{\ie}{i.e.\@ }

\newcommand{\shiftleft}[2]{\makebox[0pt][r]{\makebox[#1][l]{\rotatebox[origin=c]{90}{#2}}}}
\def\nespace{\hskip\fontdimen2\font\relax}

\begin{document}

\title{Multiplicative modulations in hue-selective cells enhance unique hue representation}

\author[1, 2, *]{Paria Mehrani}
\author[1, 2]{Andrei Mouraviev}
\author[1, 2]{John K. Tsotsos}

\affil[1]{Department of Electrical Engineering and Computer Science, York University, Toronto, Canada}
\affil[2]{The Center for Vision Research, York University, Toronto, Canada}

\affil[ ]{\vskip 0em \tt \small {paria@eecs.yorku.ca, andrei.mouraviev@gmail.com, tsotsos@eecs.yorku.ca}}

\date{}

\providecommand{\keywords}[1]{\textbf{\textit{Keywords---}} #1}

\twocolumn[
\begin{@twocolumnfalse}
	\maketitle
	\begin{abstract}
		There is still much to understand about the color processing mechanisms in the brain and the transformation from cone-opponent representations to perceptual hues. Moreover, it is unclear which areas(s) in the brain represent unique hues. We propose a hierarchical model inspired by the neuronal mechanisms in the brain for local hue representation, which reveals the contributions of each visual cortical area in hue representation. Local hue encoding is achieved through incrementally increasing processing nonlinearities beginning with cone input. Besides employing nonlinear rectifications, we propose multiplicative modulations as a form of nonlinearity. Our simulation results indicate that multiplicative modulations have significant contributions in encoding of hues along intermediate directions in the MacLeod-Boynton diagram and that model V4 neurons have the capacity to encode unique hues. Additionally, responses of our model neurons resemble those of biological color cells, suggesting that our model provides a novel formulation of the brain's color processing pathway. 
	\end{abstract}
\end{@twocolumnfalse}
\vspace{1.5cm}
]

The color processing mechanisms in the primary visual cortex and later processing stages are a target of debate among color vision researchers. What is also unclear is which brain area represents unique hues, those pure colors unmixed with other colors. In spite of a lack of agreement on color mechanisms in higher visual areas, human visual system studies confirm that color encoding starts with three types of cones forming the LMS color space. Cones send opponent feedforward signals to LGN cells with single-opponent receptive fields \cite{Reid2002}. Cone-opponent mechanisms such as those in LGN were the basis of the ``Opponent Process Theory'' of Hering~\cite{Hering}, where he introduced unique hues. The four unique hues, red, green, yellow and blue were believed to be encoded by cone-opponent processes. Later studies~\cite{Derrington1984chromatic,DeValois2000physio,Sun2006ConeInput}, however, confirmed that the cone-opponent mechanisms of earlier processing stages do not correspond to Hering's red vs. green and yellow vs. blue opponent processes. In fact, they observed that the color coding in early stages is organized along the two dimensions of the MacLeod and Boynton (MB) ~\cite{Macleod_Boyton} diagram.\nespace That is, along L\nespace vs.\nespace M and S\nespace vs.\nespace LM axes\footnote{In the MB diagram, horizontal and vertical axes represent excitations of cones in an equi-luminance plane. The horizontal axis, often referred to as the L\nespace vs.\nespace M axis, corresponds to opposing signal from L and M cones. The vertical axis, referred to as S\nespace vs.\nespace LM, represents the opposing signal from S cones against the combination of signals from L and M cones.}. 

Beyond LGN, studies on multiple regions in the ventral stream show an increase in nonlinearity with respect to the three cone types from LGN to higher brain areas~\cite{Hanazawa} and also a shift of selectivity toward intermediate hues in the MB diagram with more diverse selectivity in later processing stages \cite{Kuriki2015fMRIhue}. Specifically, in V1, some suggested that neurons representing local hue have single-opponent receptive fields similar to LGN cells~\cite{johnson2004cone,shapley2002neural} with comparable chromatic selectivities obtained by a rectified sum of the three cone types \cite{Lennie1990chromatic} or by combining LGN activations in a nonlinear fashion \cite{DeValois2000physio}.  In contrast, Wachtler~\etal \cite{Wachtler2003} found that the tunings of V1 neurons are different from LGN and that the responses in V1 are affected by context. 

Although Namima \etal \cite{Namima2014} found neurons in V4, AIT and PIT to be luminance-dependent and that the effect of luminance in responses of neurons varies from one stimulus color to another, others reported that in the Macaque extrastriate cortex, millimeter-sized neuron modules called globs, have  luminance-invariant color tunings~\cite{Conway2007}. Within globs in V4, clusters of hue-selective patches with sequential representation following the color order in the HSL space were identified, which were called ``rainbows of patches'' \cite{li2014Map} (See Figure \ref{subFig:colorMap_clusters} as an example)).  A similar observation was noted by Conway and Tsao \cite{Conway_Tsao2009}, who suggested that cells in a glob are clustered by color preference and form the hypothesized color columns of Barlow~\cite{Barlow1986}. Patches in each cluster have the same visual field location with a great overlap in their visual field with their neighboring patches \cite{li2014Map, Conway2007}. Moreover, each color activates 1-4 overlapping patches and neighboring patches are activated for similar hues. Comparable findings in V2 were reported in~\cite{XiaoTopoV2}. Following these observations, Li \etal~\cite{li2014Map} suggested that different multi-patch patterns represent different hues, and such a distributed and combinatorial color representation could encode the large space of physical colors, given the limited number of neurons in each cortical color map. Other studies also suggested that glob populations uniformly represent color space \cite{Bohon2016} with narrow tunings for glob cells \cite{Bohon2016, Schein1990_silent_surround,Zeki1980}.

Not only is there disagreement about the color processing mechanisms in the visual cortex, but also which region in the brain represents unique hues.\nespace Furthermore, transformation mechanisms from cone-opponent responses to unique hues are unclear. While unique red is found to be close to the +L axis in the MB diagram, unique green, yellow and blue hues cluster around intermediate directions \cite{Webster2000variations}, not along cone-opponent axes.\nespace Perhaps clustering of the majority of unique hues along intermediate directions could describe the suggestion by Wuerger \etal~\cite{Wuerger2005} who proposed that the encoding of unique hues, unlike the tuning of LGN neurons, needs higher order mechanisms such as a piecewise linear model in terms of cone inputs. The possibility of unique hue representations in V1 and V2 was rejected in \cite{Stoughton2008}, who like others \cite{Zeki1980, Komatsu1992} observed neurons in PIT show selectivities to all hue angles\footnote{On the color wheel, each hue can be represented using its counterclockwise angle from 3 o'clock on the circle. Similarly, in the MB space, hues can be mapped to a unit circle and represented by their angle on the circle.} and that there are more neurons selective to those close to unique hues. The choice of stimuli for recordings in~\cite{Stoughton2008} was then challenged in \cite{Mollon2009} commenting that it is still unclear whether or not unique hues are represented in IT. Similarly, Zaidi~\etal~\cite{Zaidi2014} observed no significance of unique hues in human subjects and responses of IT neurons.

Among all the attempts to understand the neural processes for transformation from cone-opponency to perceptual colors, a number of computational models tried to suggest mechanisms for this problem and other aspects of color representation in higher areas \cite{Dufort_Lumsden, Courtney1995, Wray1996, DeValois1993multi, lehky1999seeing}. These models, however, are one-layer formulations of perceptual hue encoding, or in other words, the totality of processing in these models is compressed into a single layer process. The end result may indeed provide a suitable model in the sense of its input-output characterization. However, it does not make an explicit statement about what each of the processing areas of the visual cortex are contributing to the overall result and they do not shed light upon the mystery of color representation mechanisms in the brain.

In this article, we introduce a computational color processing model that as Brown~\cite{Brown2014tale} argues, helps in ``understand[ing] how the elements of the brain work together to form functional units and ultimately generate the complex cognitive behaviors we study''. For this purpose, we build a hierarchical framework, inspired by neural mechanisms in the visual system, that explicitly models neurons in each of LGN, V1, V2, and V4 areas and reveals how each visual cortical area participates in the process. In this model, nonlinearity is gradually increased in the hierarchy as observed by~\cite{Hanazawa}. In particular, while a half-wave rectifier unit keeps the V1 tunings similar to those of LGN~\cite{Lennie1990chromatic}, it makes V1 neurons nonlinear in terms of cone inputs. In V2, in addition to single-opponent cells, we propose employing neurons with multiplicative modulations, which not only introduce another form of nonlinearity but also allow neuronal interactions in the form of mixing of color channels as well as a decrease in the tuning bandwidths. De Valois et al. \cite{DeValois1993multi} suggested that additive or subtractive modulation of cone-opponent cells with S-opponent cell responses rotates the cone-opponent axes to red-green and blue-yellow directions. Here, we achieved this rotation with multiplicative modulations of V1 L- and M-opponent cell activations with V1 S-opponent neuron responses. We call these cells ``multiplicative V2'' neurons. Finally, V4 responses are computed by linearly combining V2 activations with weights determined according to tuning peak distances of V2 cells to the desired V4 neuron tuning peak. 

\begin{figure*}[p]
	\centering
	\subfigure[]{\includegraphics[width = 0.8\textwidth]{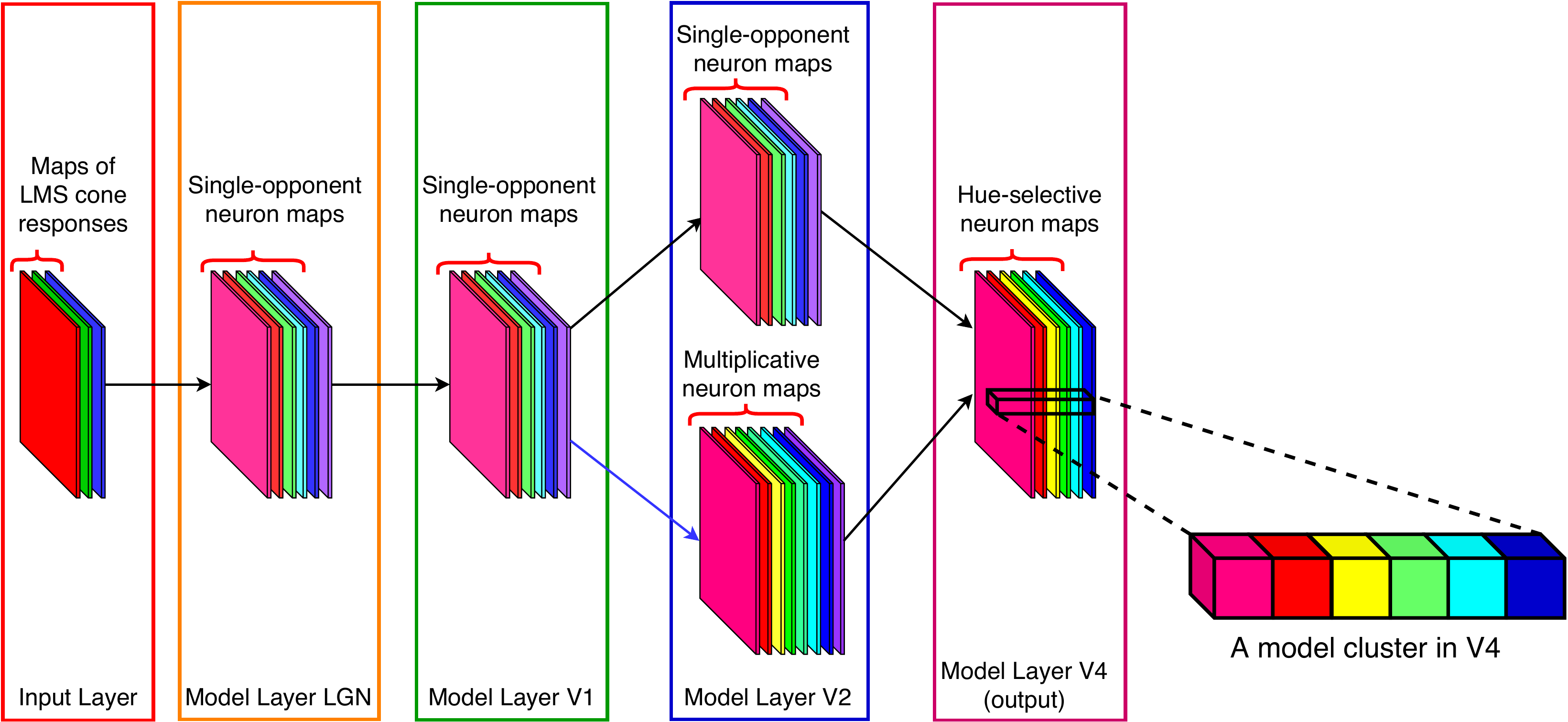}\label{fig:network}}

	\subfigure[]{\includegraphics[width=0.8\textwidth]{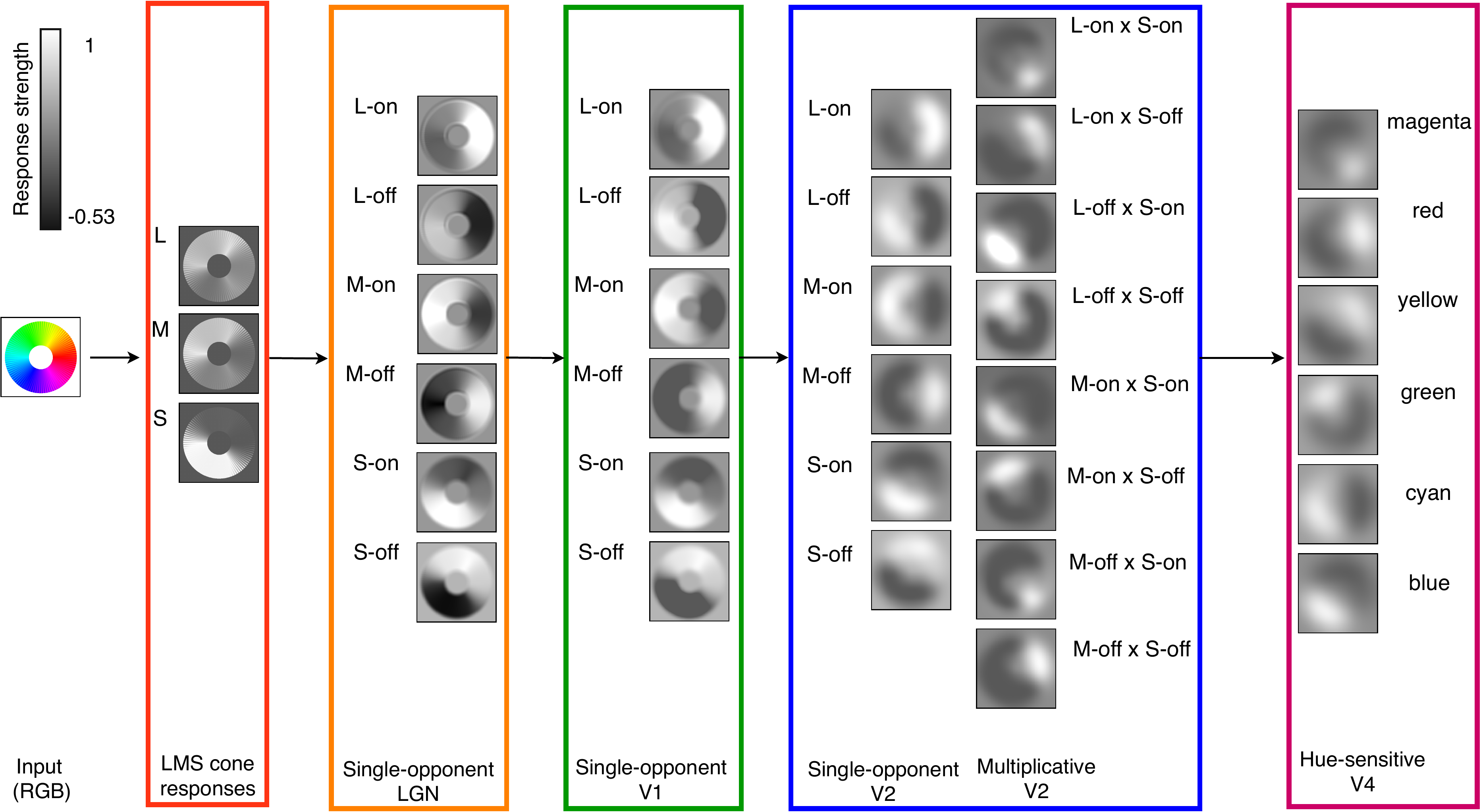}	\label{fig:qual_examples}}
	\caption{\subref{fig:network} An illustration of the proposed hierarchical model for local hue representation (best seen in color). Each layer of this model implements neurons in a single brain area. Each map within a layer consists of neurons of a single type with receptive fields spanning the visual field of the model. For example, a map of neurons selective to red hue in model layer V4. The leftmost layer in this figure shows the input to the network with the LMS cone activations. A combination of cone responses with opposite signs results in neurons with single-opponent receptive fields in model layers LGN, V1, and V2. Multiplicatively modulated V2 cell activations are the result of V1 L- and M-opponent cell activations multiplied with V1 S-opponent neuron responses. The output layer of the model consists of hue-selective V4 neurons. Note that the color employed for each feature map here is figurative and not a true representation of the hue-selectivity of its comprising neurons. In model layer V4, an example of a model cluster is shown in a larger view, similar to the clusters found in monkey V4 \cite{li2014Map}. Each model cluster corresponds to a column of the three dimensional matrix obtained by stacking V4 maps. Each element of a model cluster is called a model patch. \subref{fig:qual_examples} An example showing each layer of the hierarchical color model on an image of a hue wheel. Each layer of the network is shown by a bounding box, with a number of neuronal maps inside the box. Next to each map, the selectivity of its neurons is written. The receptive field of each neuron in these maps is centered at the corresponding pixel location. The neuron responses are shown in grayscale, with a minimum response as black, and maximum activation as white. For example, in the map for neuron type L-on in the LGN layer, strong activities are observed for neurons with receptive fields around the red hue region. The dark border around each feature map is shown only for the purpose of this figure and is not part of the activity map.}
\end{figure*}

Figure \ref{fig:network} depicts our proposed model and Figure \ref{fig:qual_examples} demonstrates our network in action. Each layer of this model implements neurons in a single brain area. Each map within a layer consist of neurons of a single type with receptive fields spanning the visual field of the model, for example, a map of neurons selective to red hue in model layer V4. The leftmost layer in this figure shows the input to the network with the LMS cone activations. 
We found that the tuning peak of multiplicatively modulated V2 cells shifts toward hues along intermediate directions in the MB space. Consequently, these neurons have substantial input weights compared to single-opponent V2 cells to V4 neurons selective to hues along intermediate directions. Moreover, we observed a gradual decrease in distance of tuning peaks to unique hue angles reported by \cite{Miyahara2003focal} from our model LGN cells to V4 neurons. Our simulation results demonstrate that responses of our network neurons resemble those of biological color cells. 

In what follows, we will make a distinction between our model and brain areas by referring to those as layers and areas, respectively. That is, a set of model neurons implementing cells in a brain area will be referred to as a model layer. For example, our model layer V2 implements cells in brain area V2.

\section*{Results}
In this section, we explain our simulation experiments, designed to make two important aspects of our model clear: 
\begin{enumerate}
	\item on the single cell level, our model neurons perform similarly to their biological counterparts.
	\item on the system level, our hierarchical model with a gradual increase in nonlinearity makes it possible to model neurons with narrow bandwidths, represent hues in intermediate directions, and represent unique hues in our model output layer V4.
\end{enumerate}
As a result, our experiments bridge a mixture of single cell examinations to evaluations of the hierarchical model as a whole. 
\vspace{-0.3cm}

\subsection*{Model Neuron Tunings}
\begin{figure*}[!p]
	\renewcommand*{\thesubfigure}{(\arabic{subfigure})}
	\setcounter{subfigure}{0}
	\centering
	\subfigure[L-on]{\makebox[20pt]{\raisebox{25pt}{\rotatebox[origin=c]{90}{LGN}}}~\includegraphics[width=0.15\textwidth]{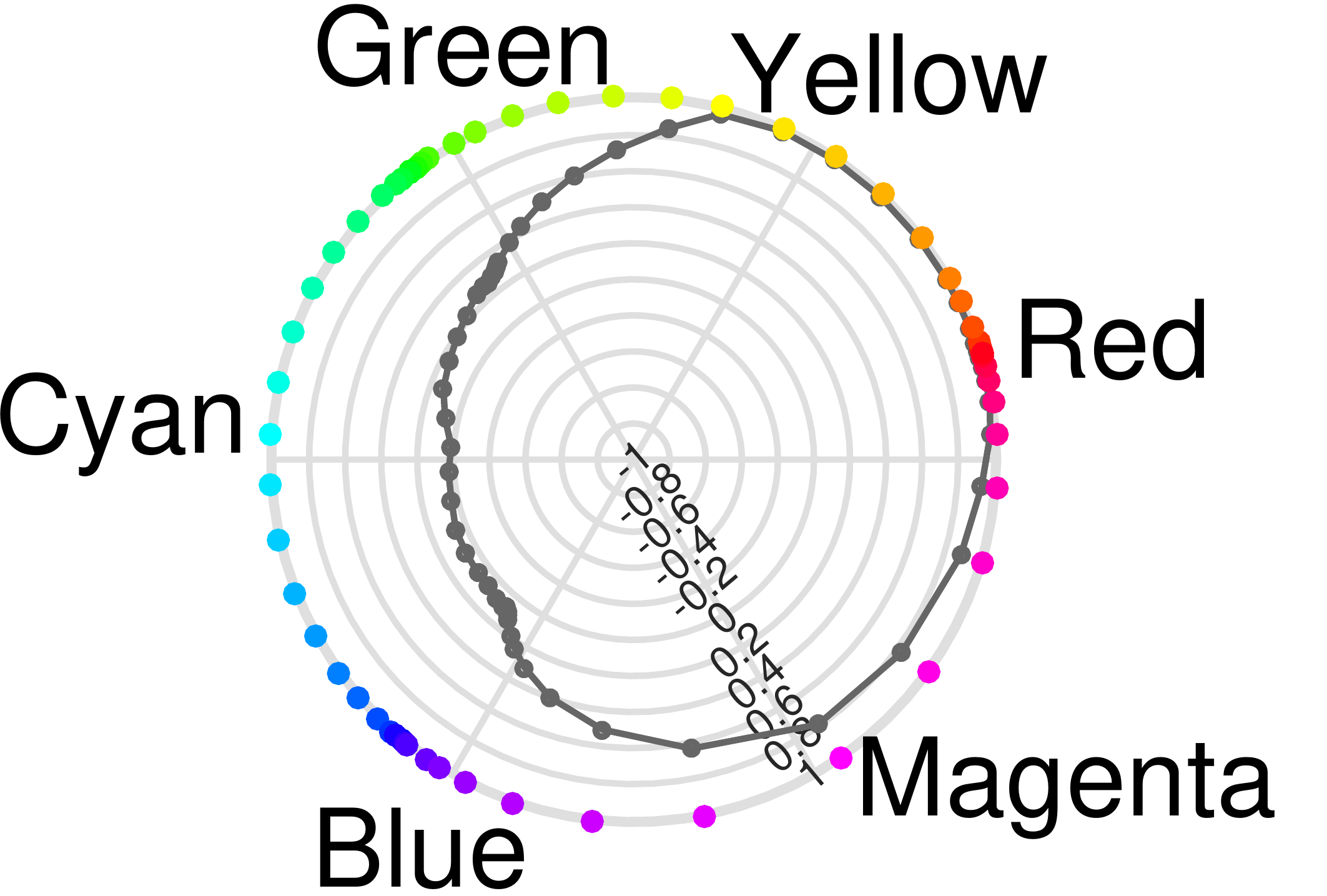}}~
	\subfigure[L-off]{\includegraphics[width=0.15\textwidth]{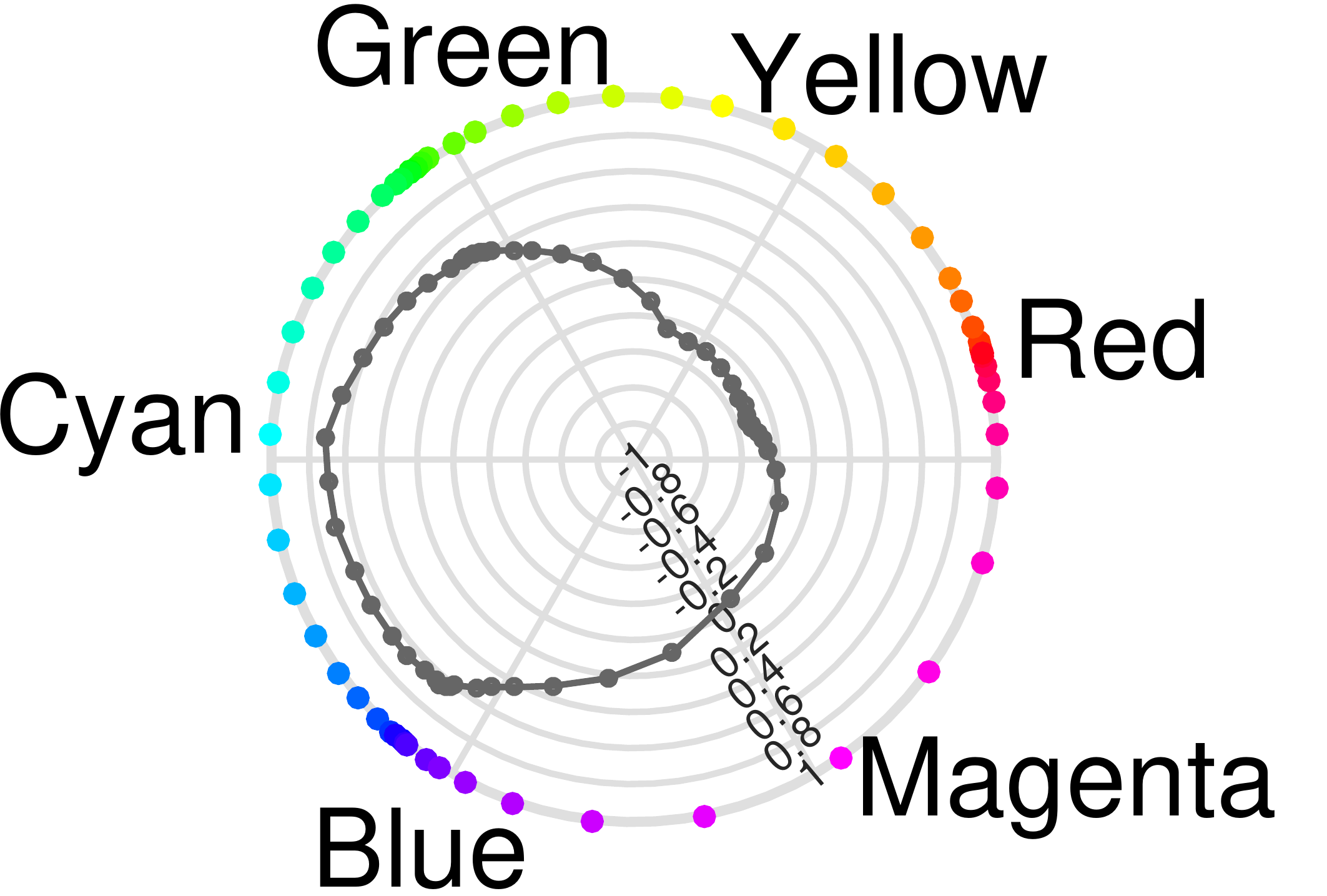}}~
	\subfigure[M-on]{\includegraphics[width=0.15\textwidth]{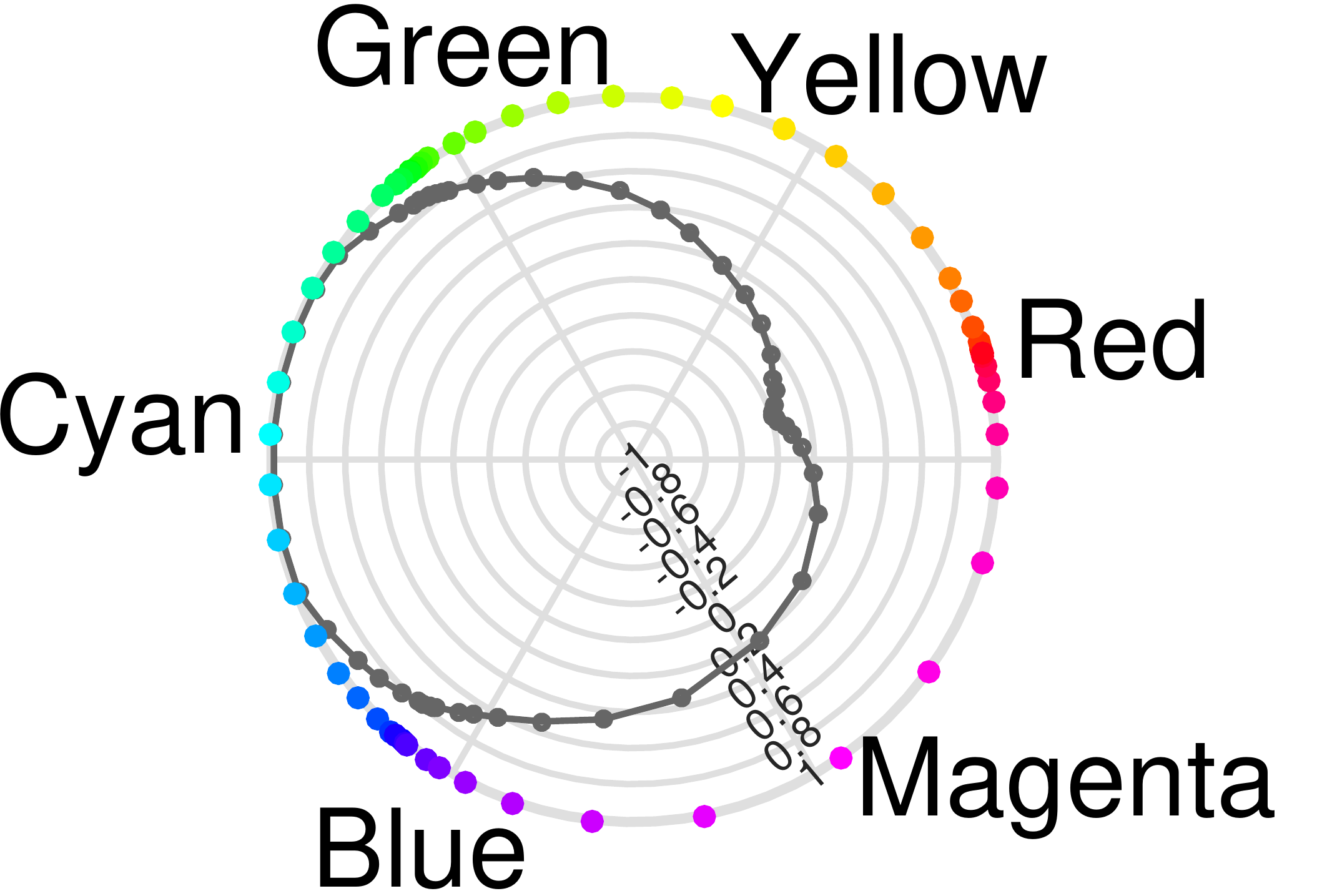}}~
	\subfigure[M-off]{\includegraphics[width=0.15\textwidth]{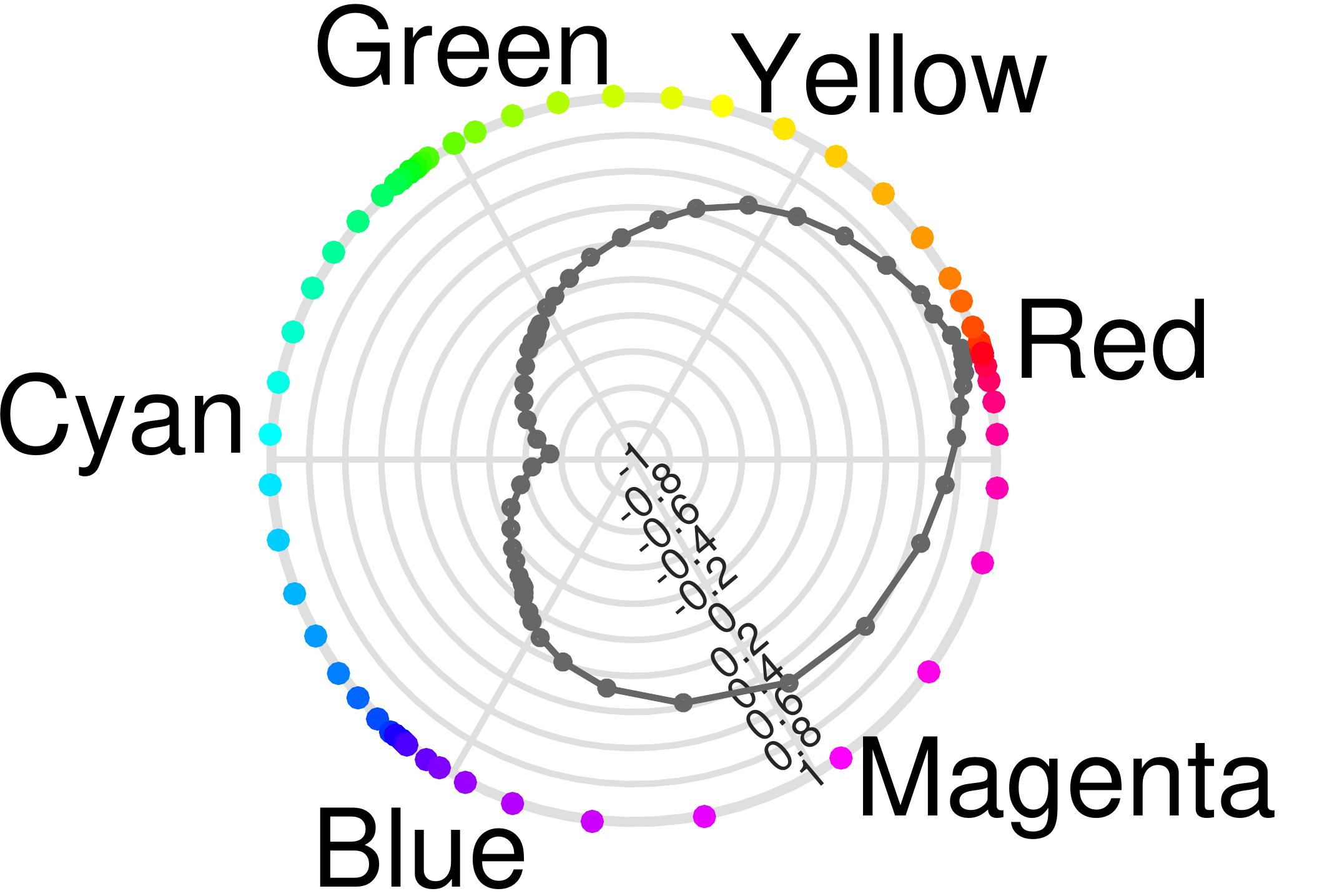}}~
	\subfigure[S-on]{\includegraphics[width=0.15\textwidth]{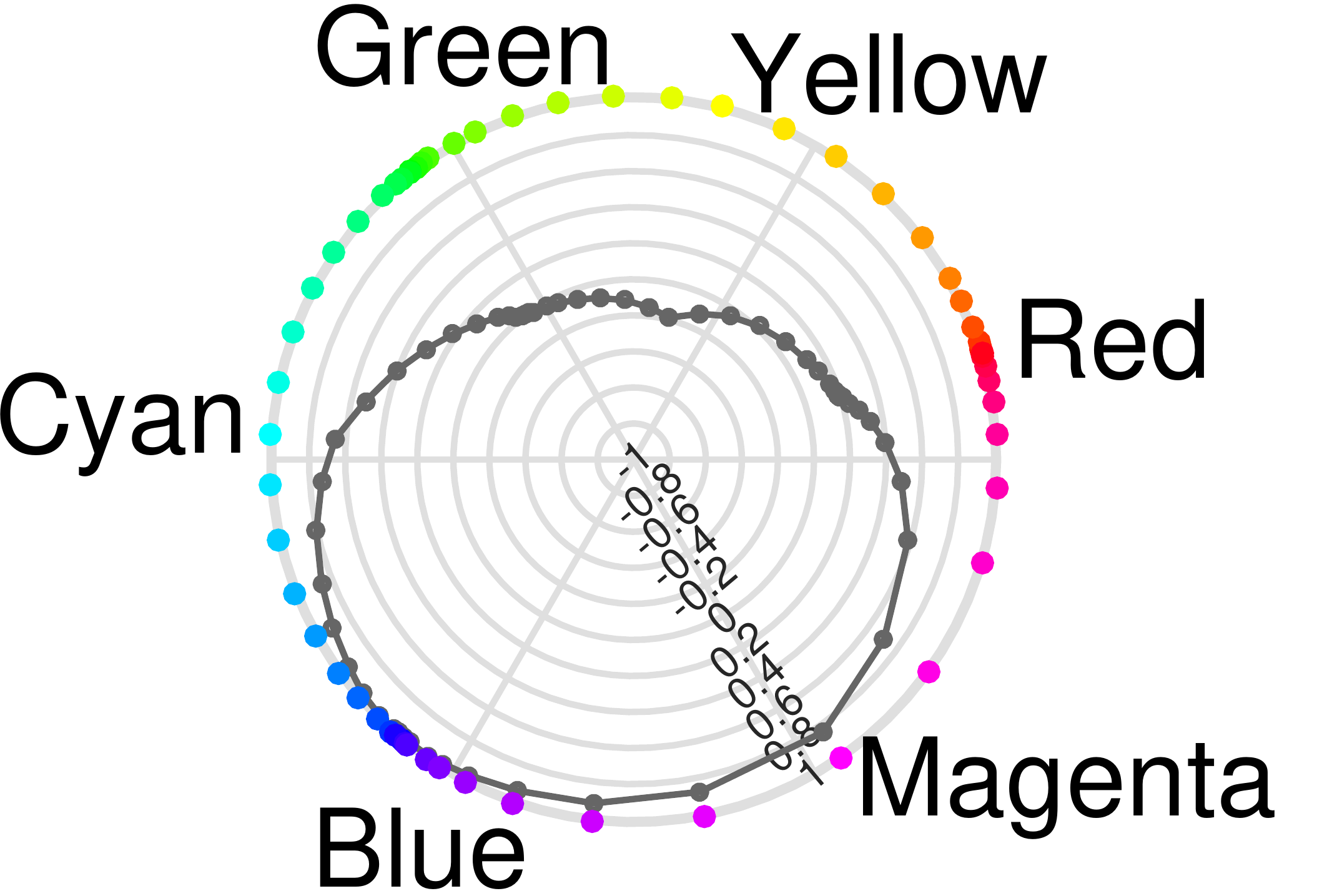}}~
	\subfigure[S-off]{\includegraphics[width=0.15\textwidth]{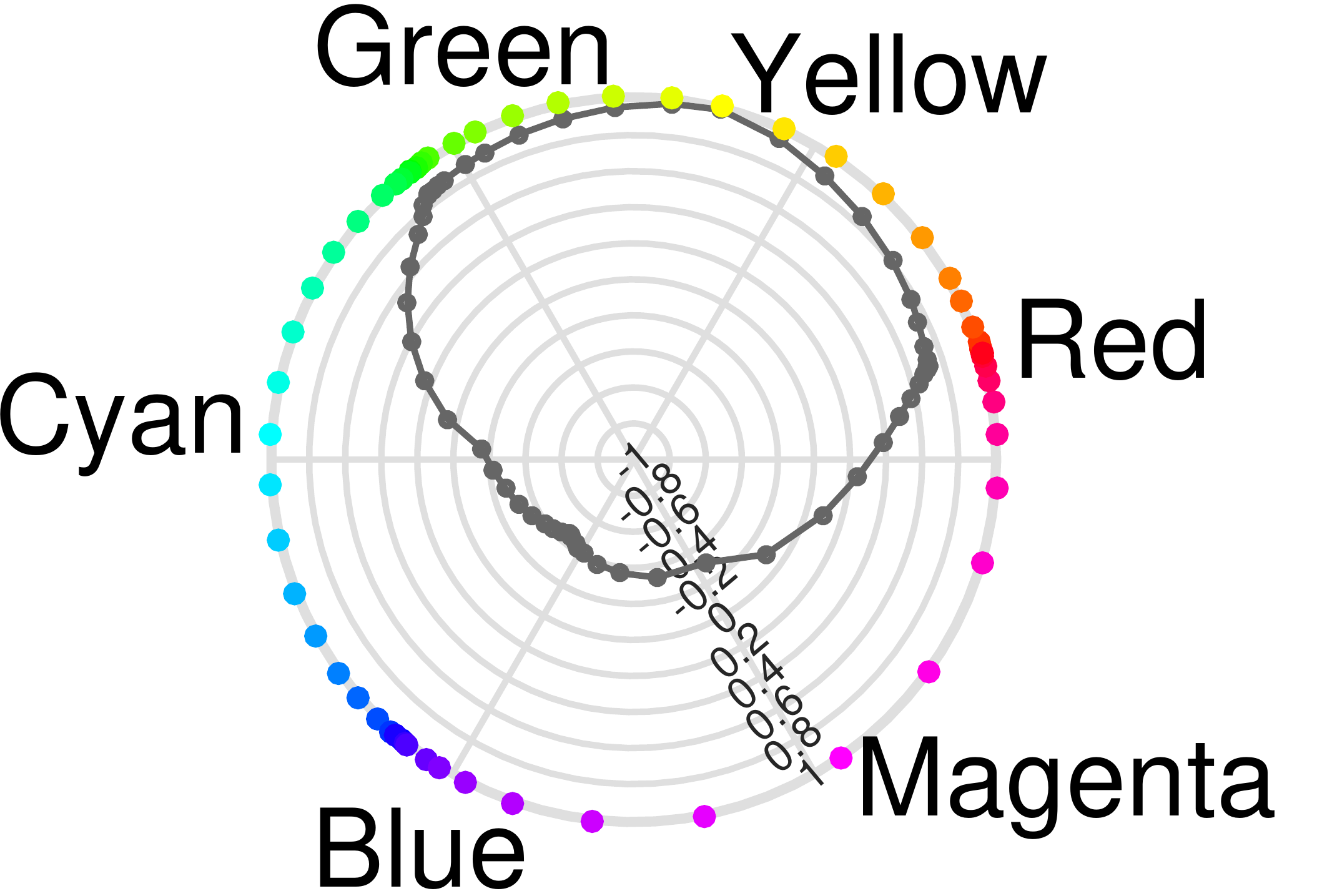}}\\ 
	\vspace{20pt}
	\subfigure[L-on]{\makebox[20pt]{\raisebox{25pt}{\rotatebox[origin=c]{90}{V1}}}~\includegraphics[width=0.15\textwidth]{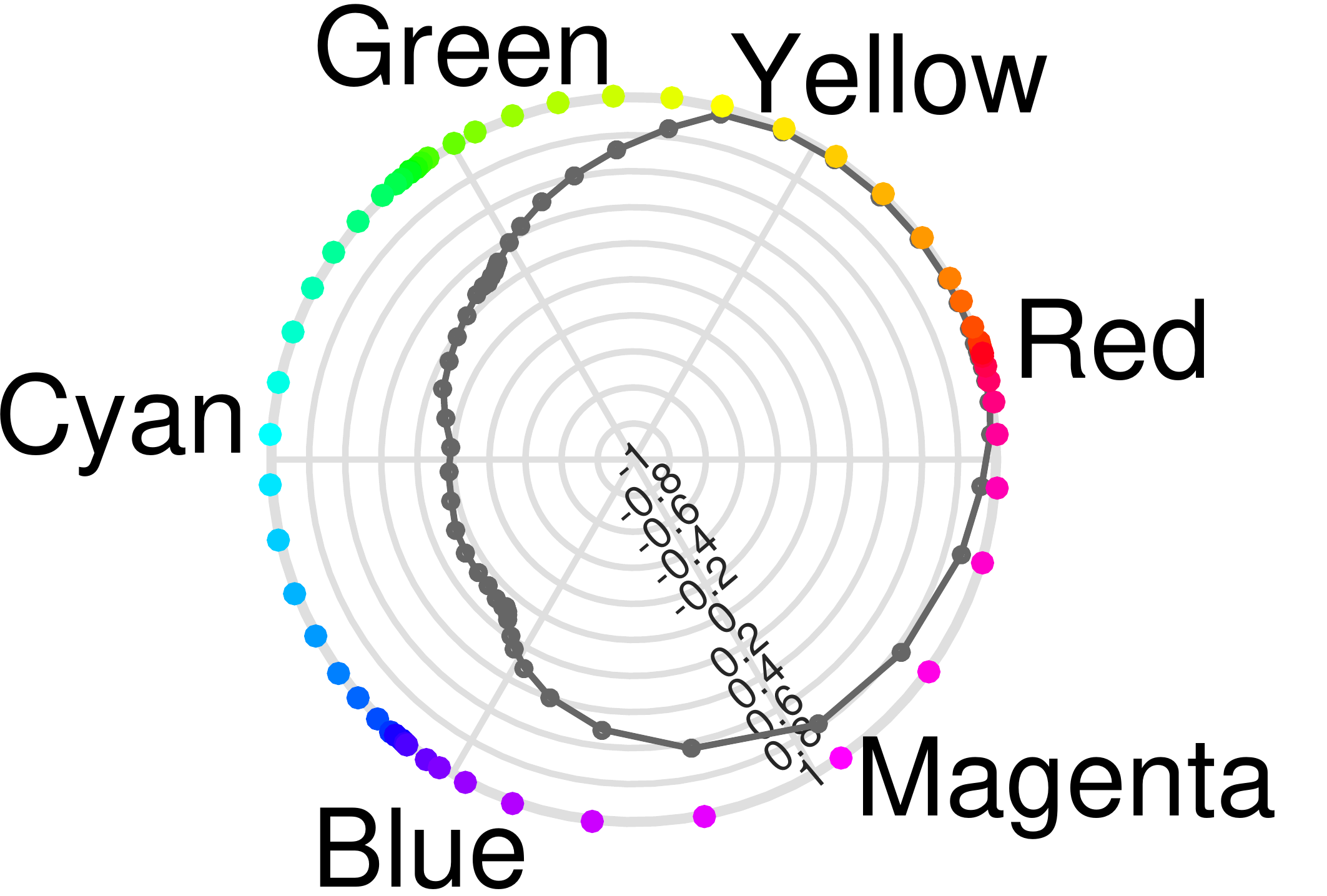}}~
	\subfigure[L-off]{\includegraphics[width=0.15\textwidth]{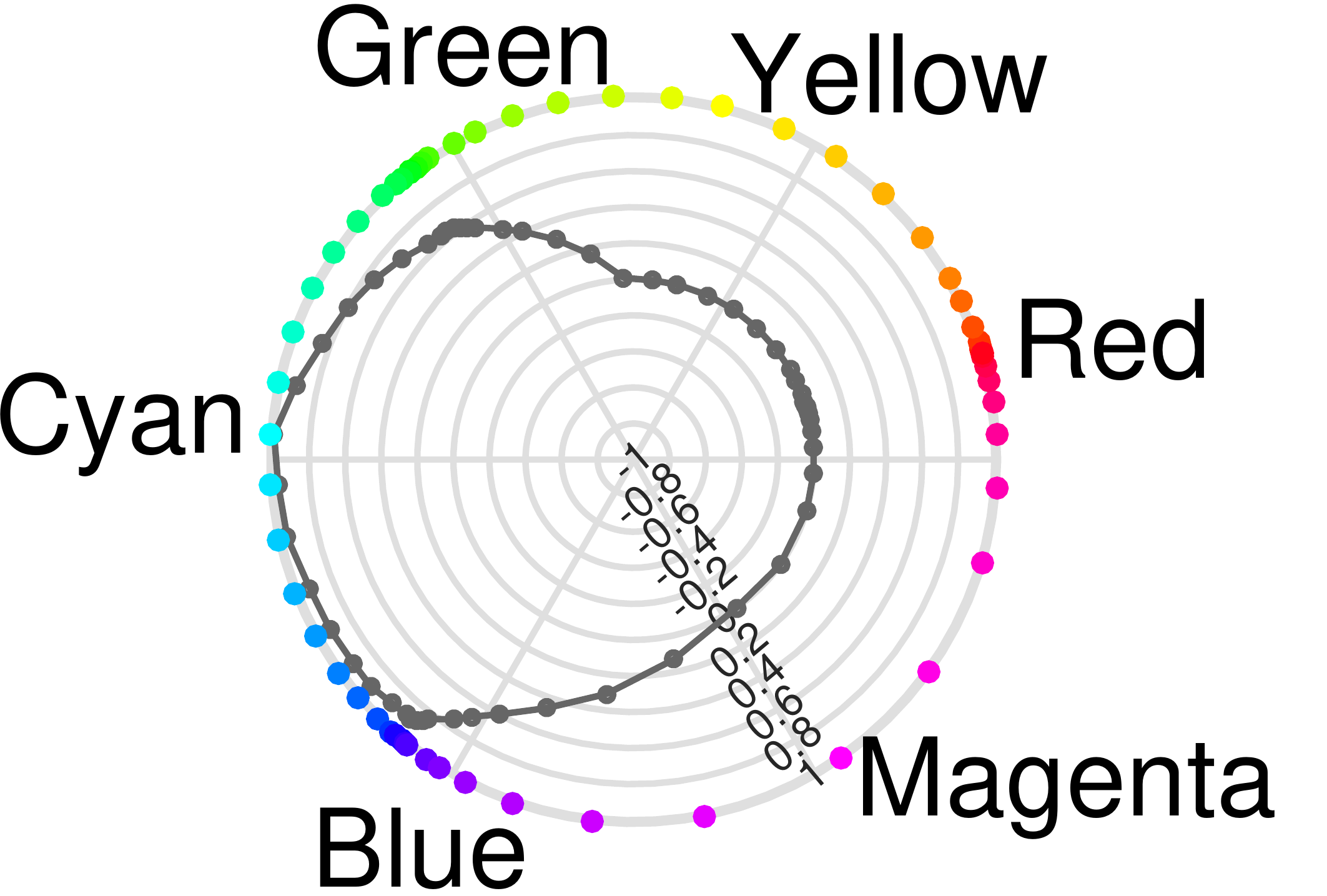}}~
	\subfigure[M-on]{\includegraphics[width=0.15\textwidth]{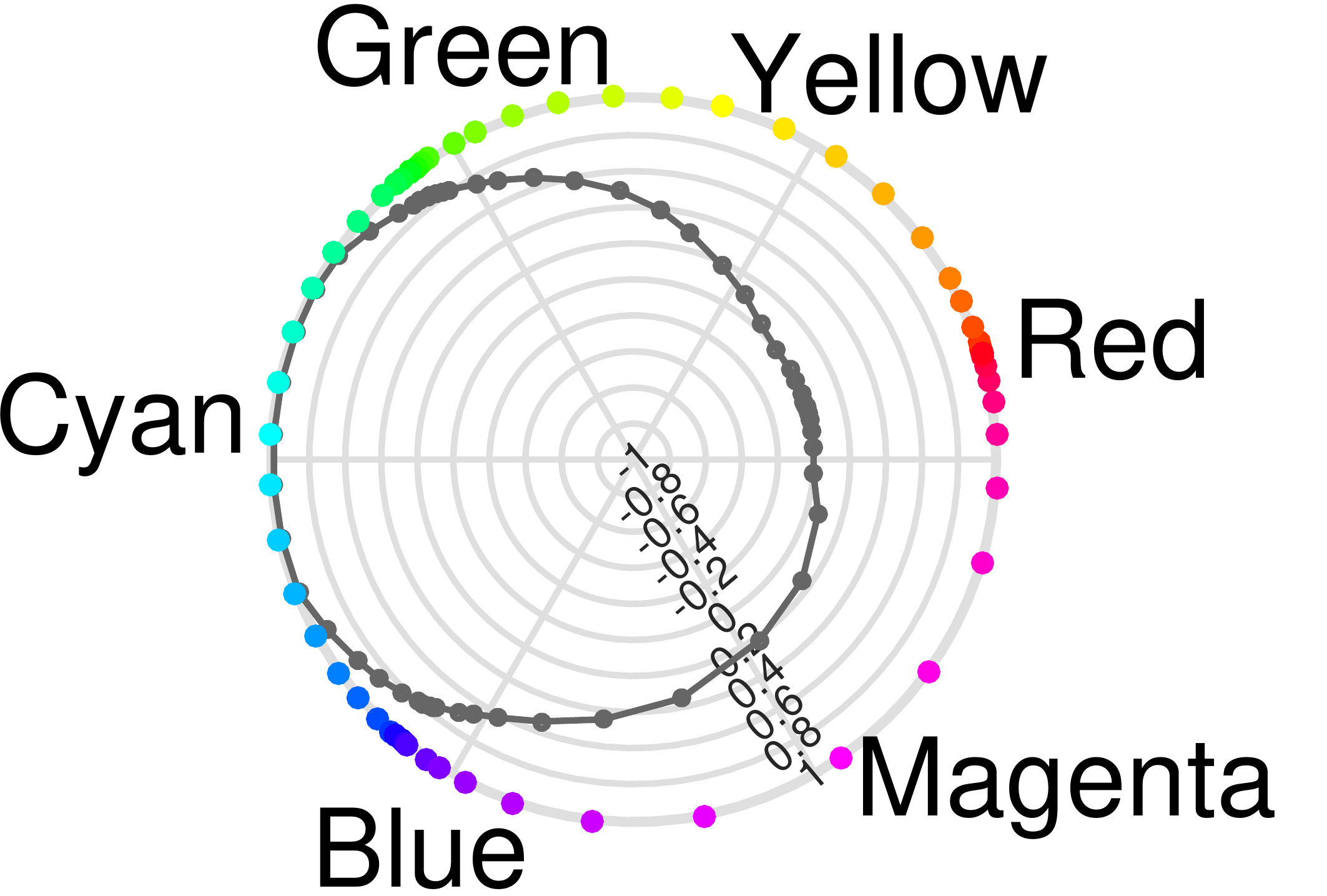}}~
	\subfigure[M-off]{\includegraphics[width=0.15\textwidth]{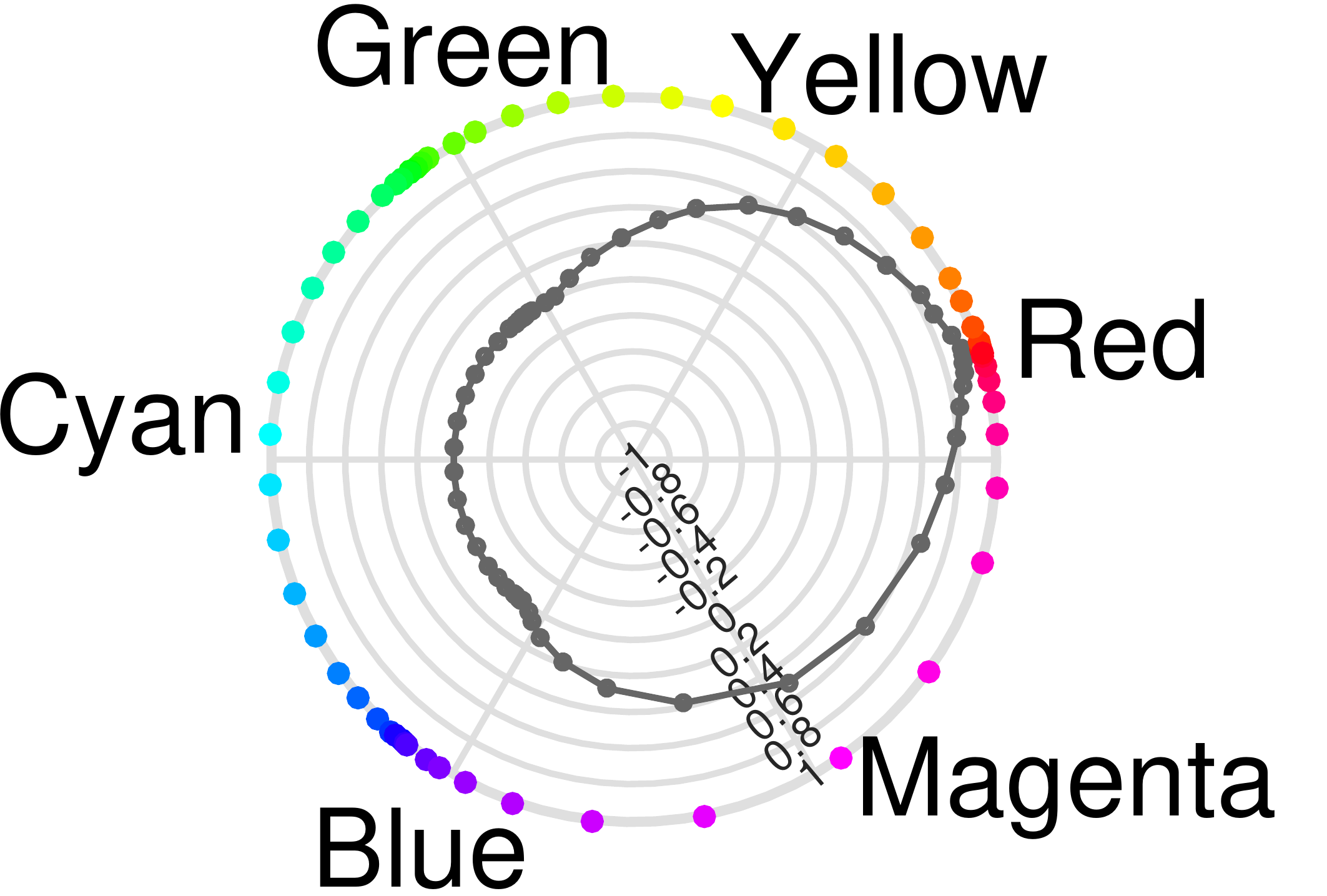}}~
	\subfigure[S-on]{\includegraphics[width=0.15\textwidth]{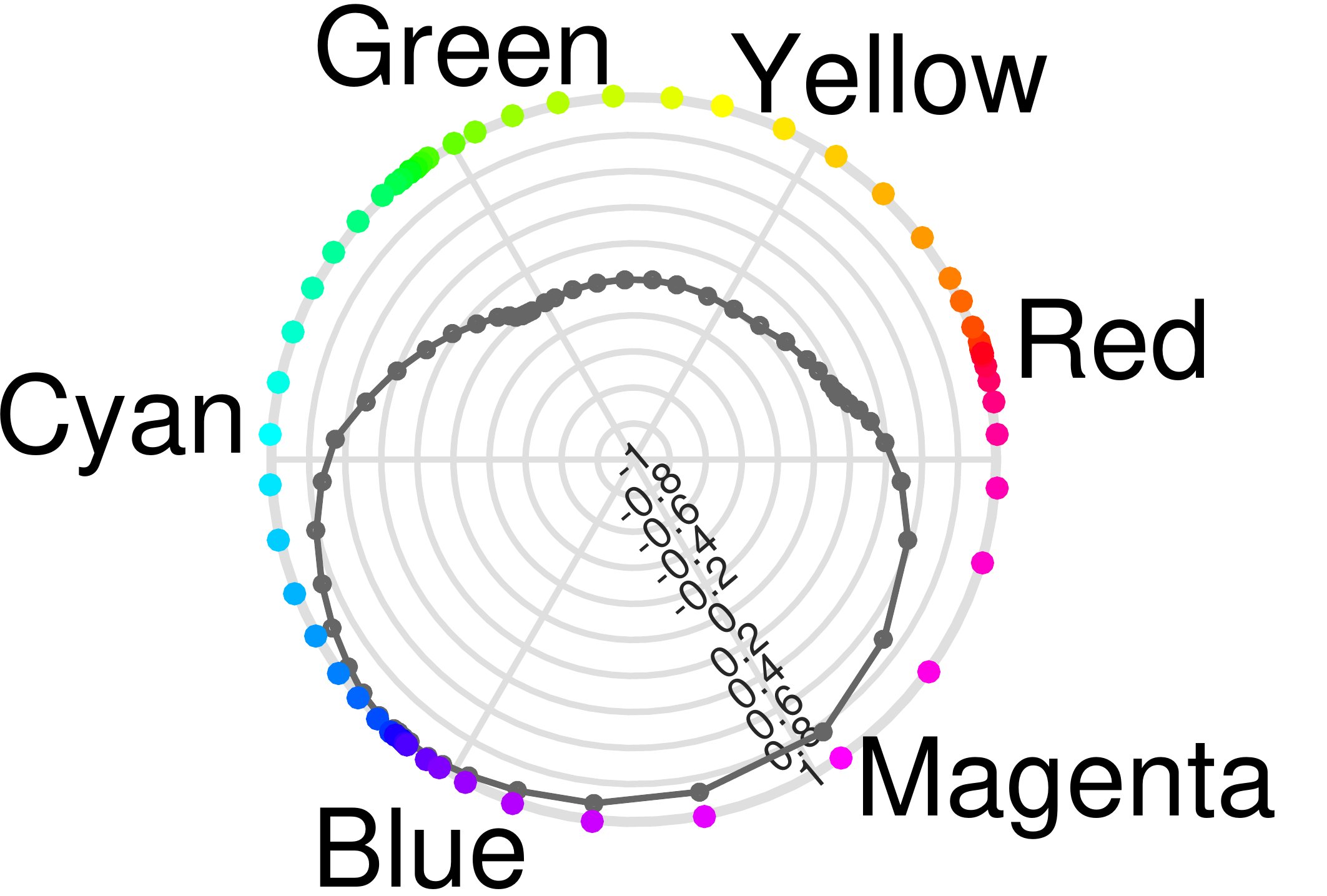}}~
	\subfigure[S-off]{\includegraphics[width=0.15\textwidth]{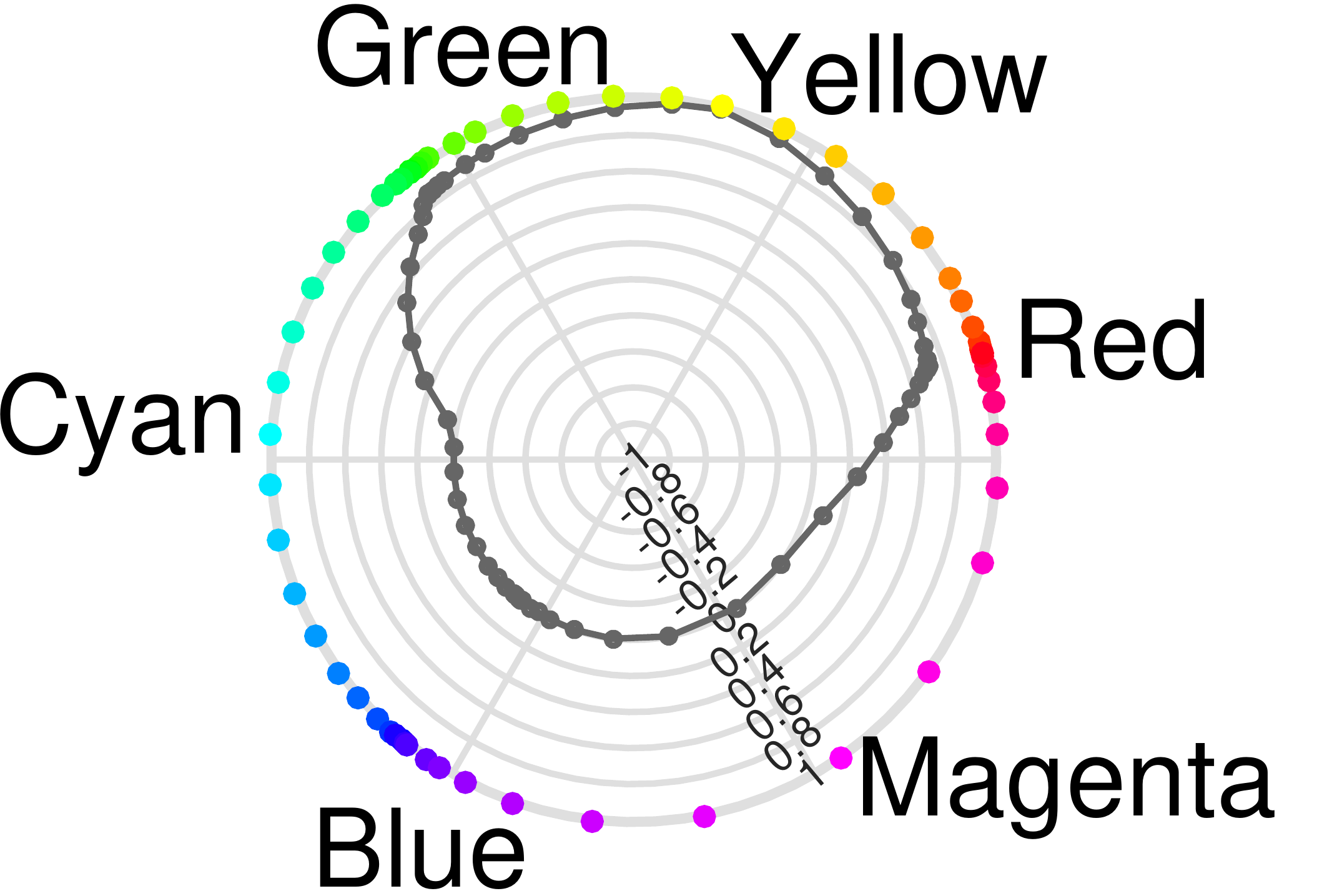}}\\
	\vspace{20pt}
	\makebox[20pt]{\raisebox{15pt}{\rotatebox[origin=c]{90}{Single-opponent V2}}}~
	\subfigure[L-on]{\includegraphics[width=0.15\textwidth]{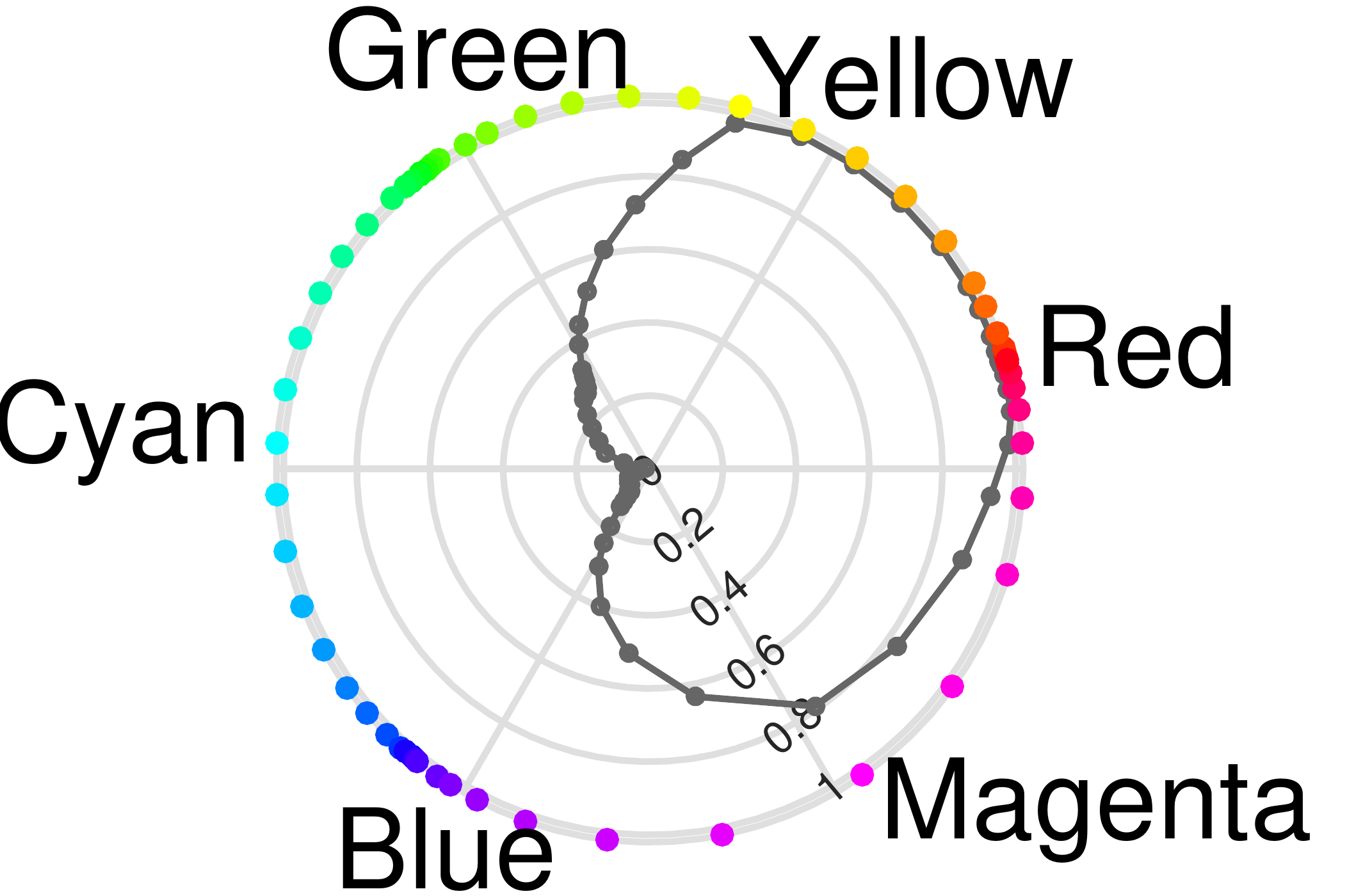}}~
	\subfigure[L-off]{\includegraphics[width=0.15\textwidth]{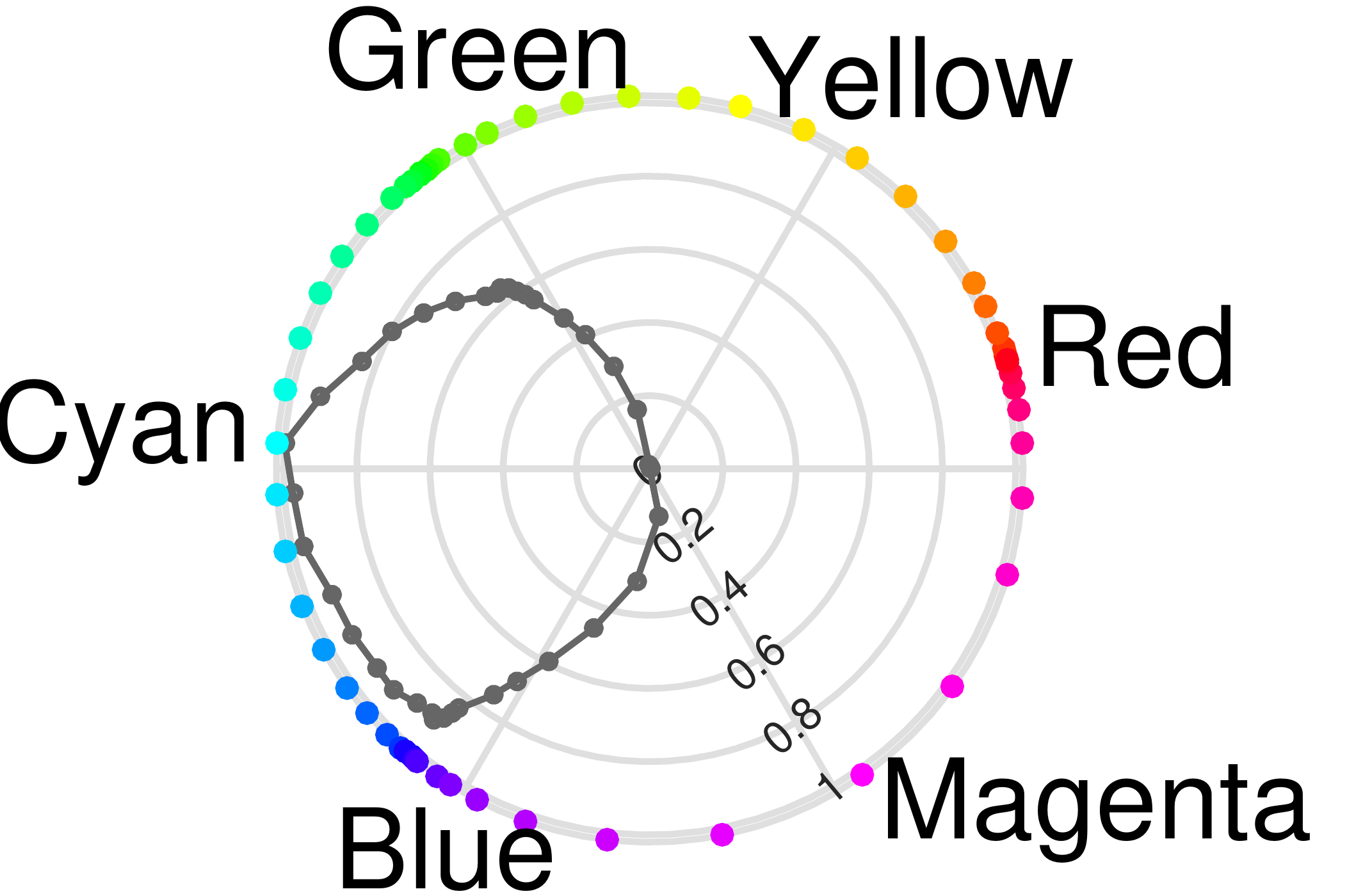}}~
	\subfigure[M-on]{\includegraphics[width=0.15\textwidth]{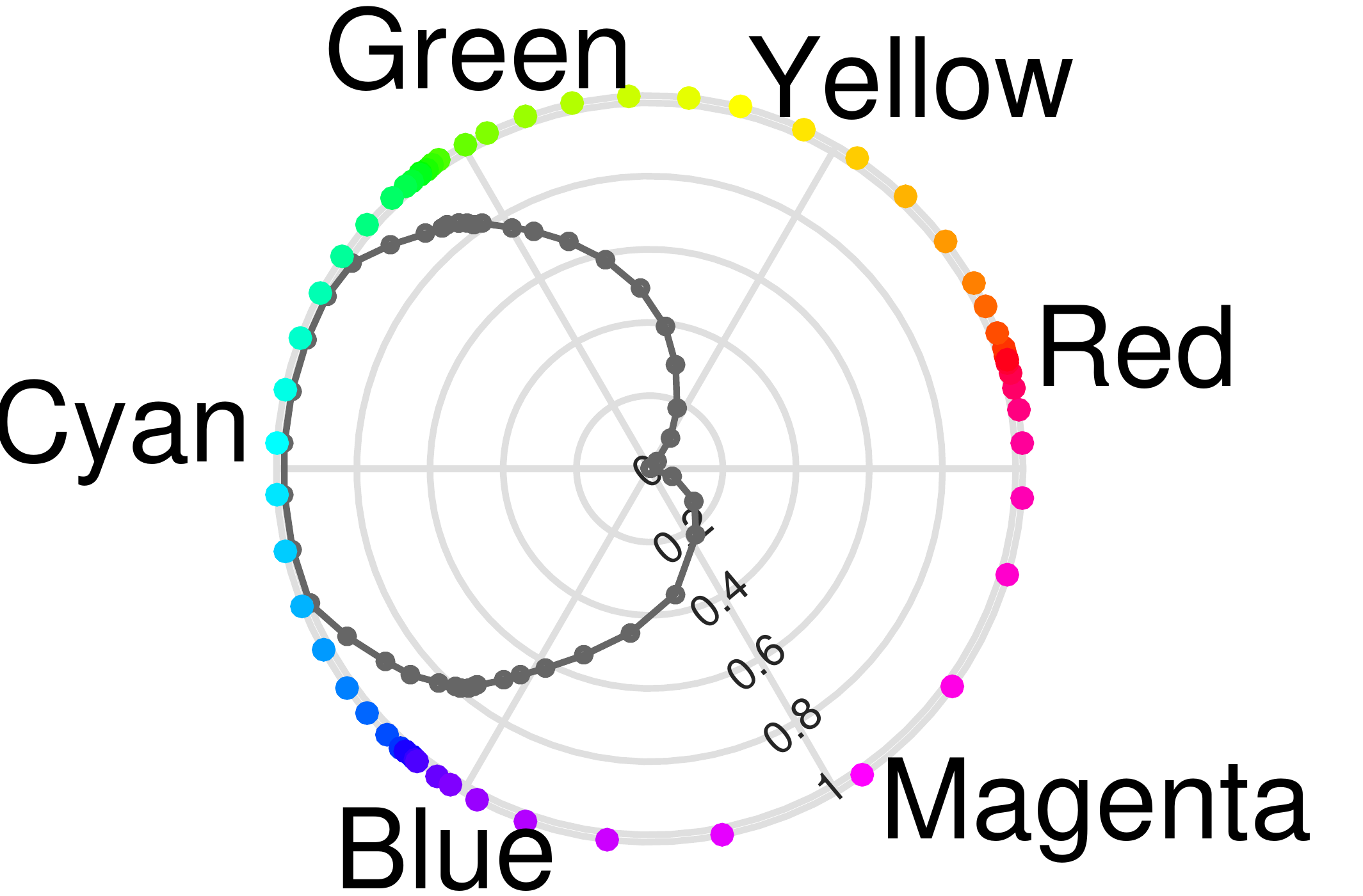}}~
	\subfigure[M-off]{\includegraphics[width=0.15\textwidth]{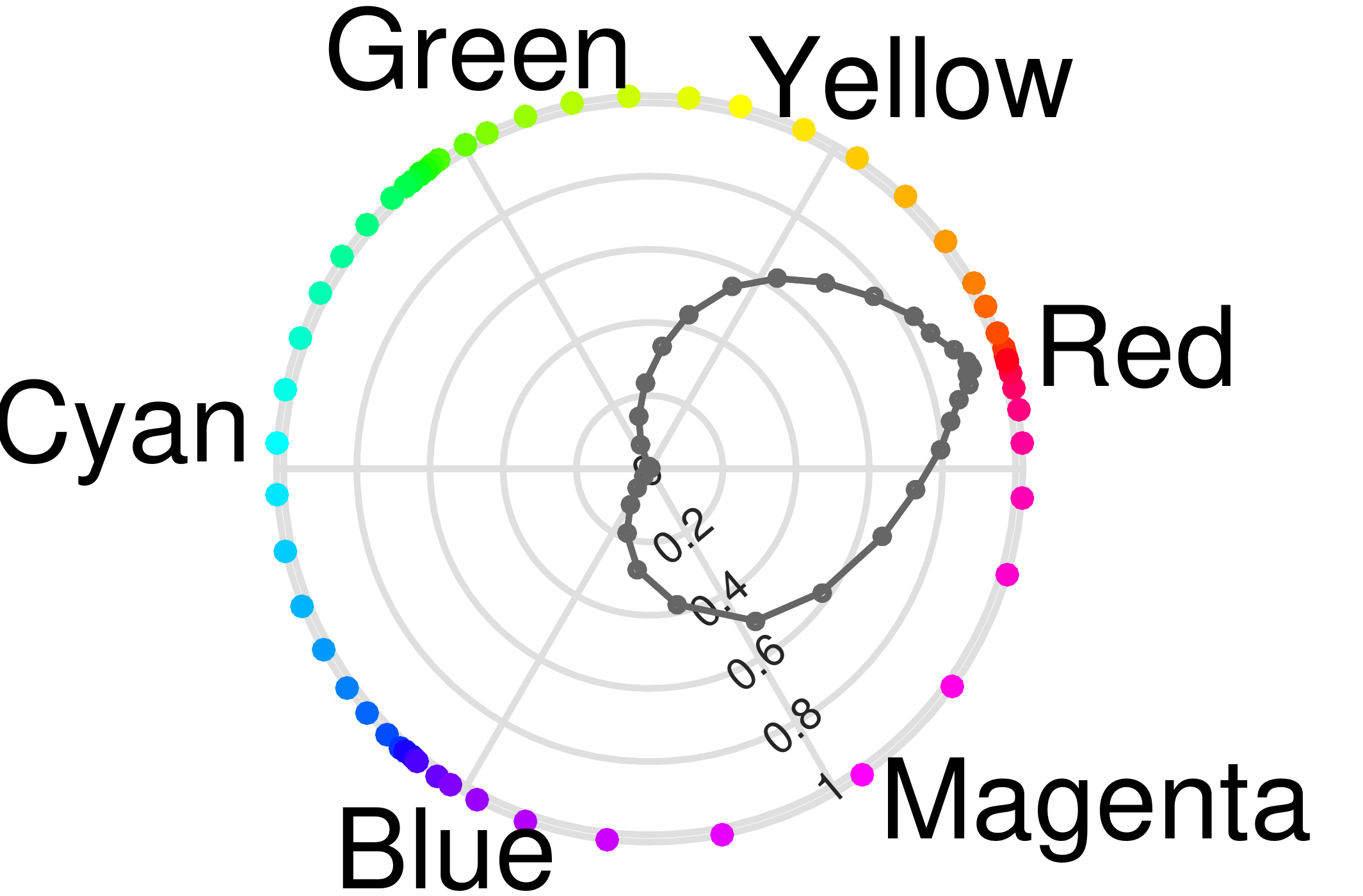}}~
	\subfigure[S-on]{\includegraphics[width=0.15\textwidth]{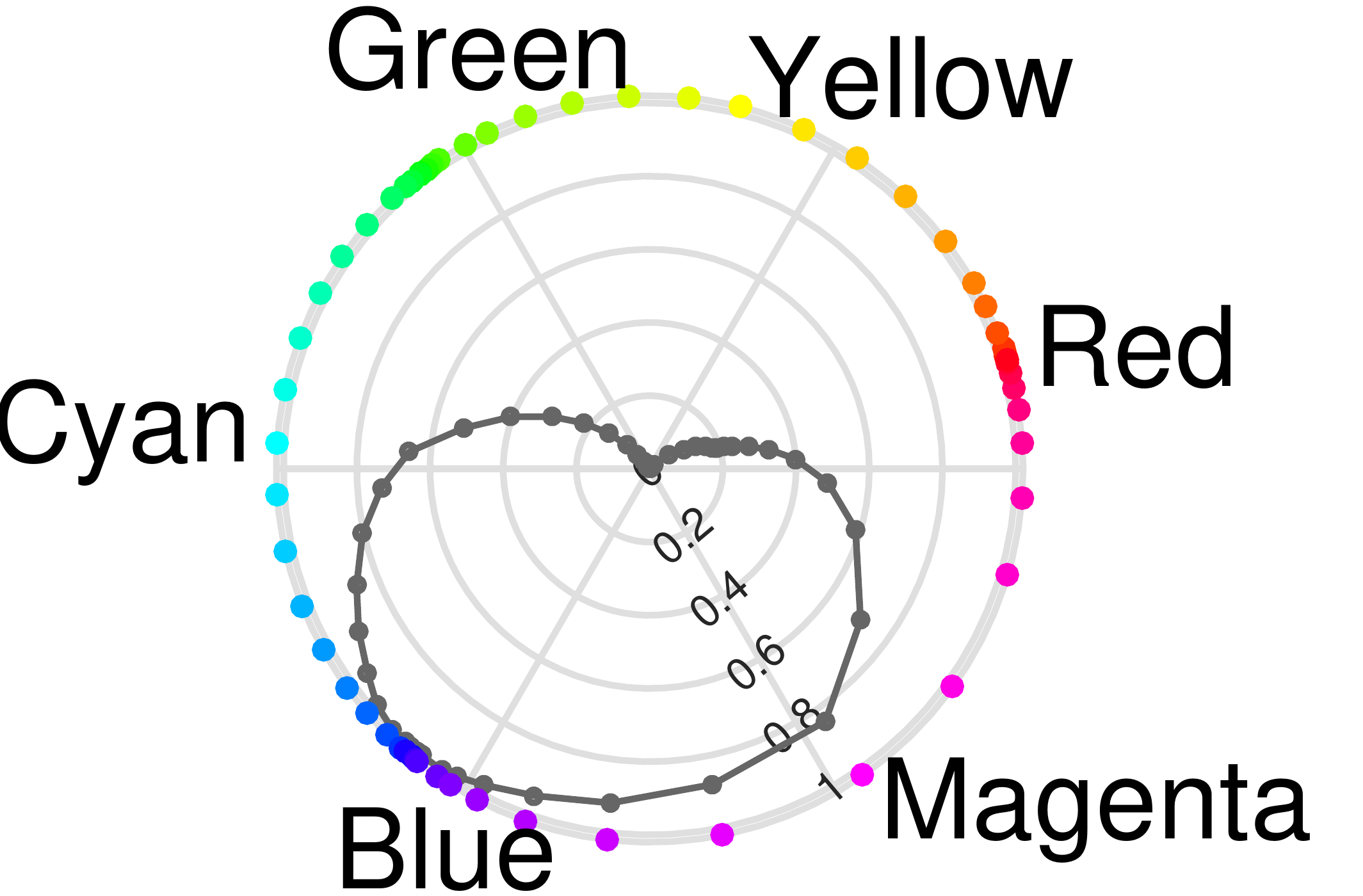}}~
	\subfigure[S-off]{\includegraphics[width=0.15\textwidth]{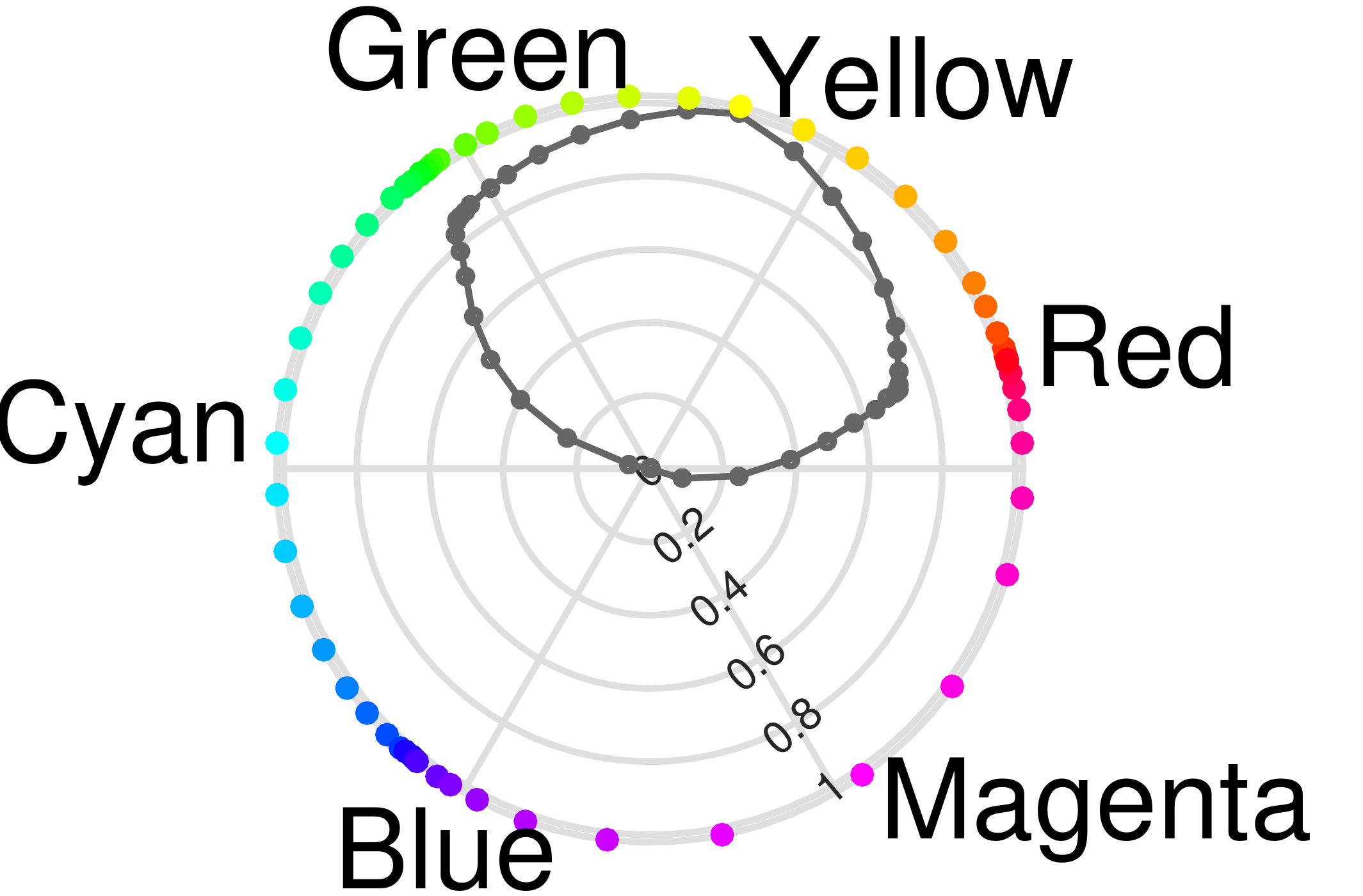}}\\
	\vspace{20pt}
	\makebox[20pt]{\raisebox{15pt}{\rotatebox[origin=c]{90}{Multiplicative V2}}}~
	\subfigure[L-on $\times$ S-on]{\includegraphics[width=0.15\textwidth]{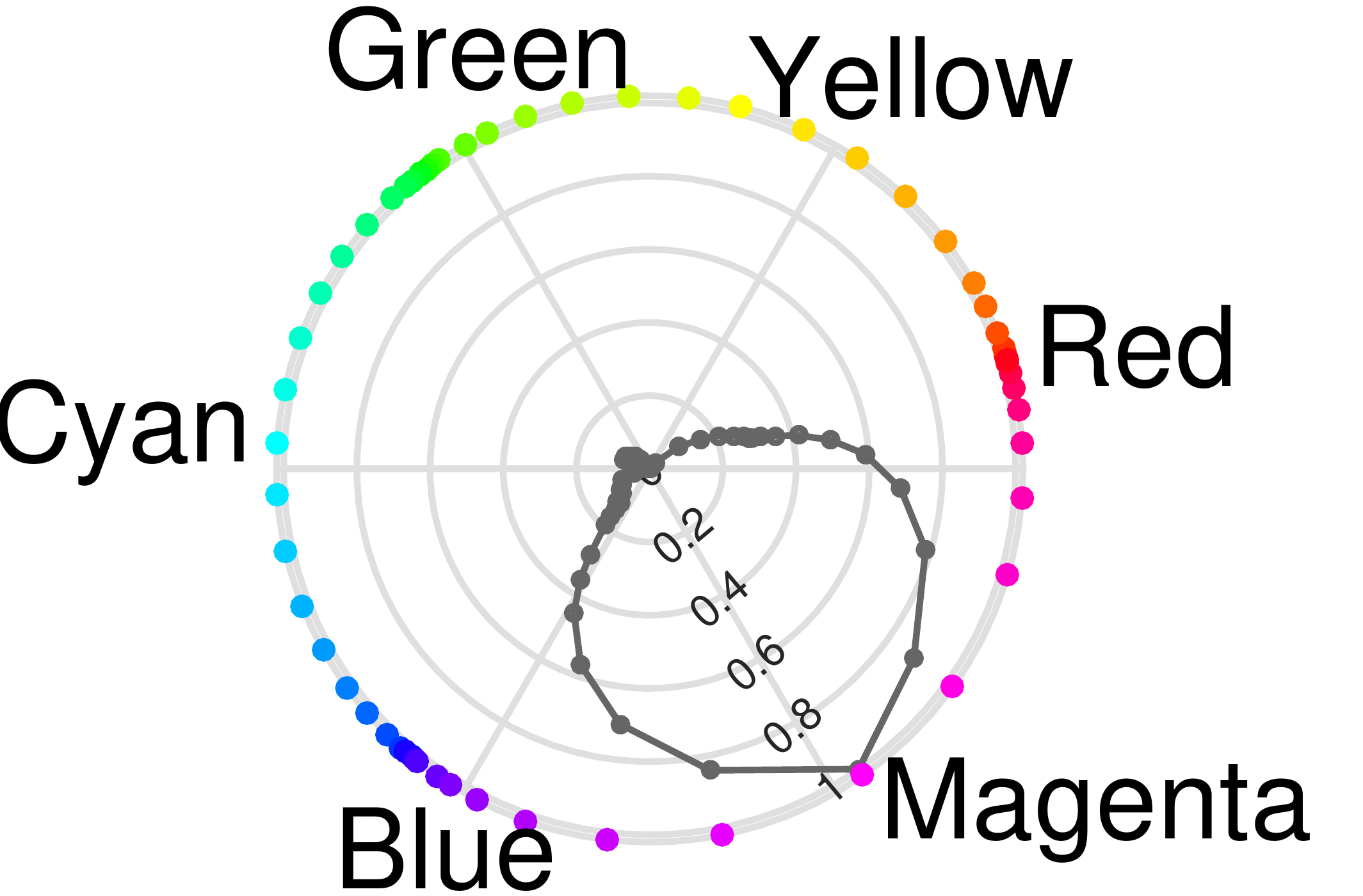}}~
	\subfigure[L-on $\times$ S-off]{\includegraphics[width=0.15\textwidth]{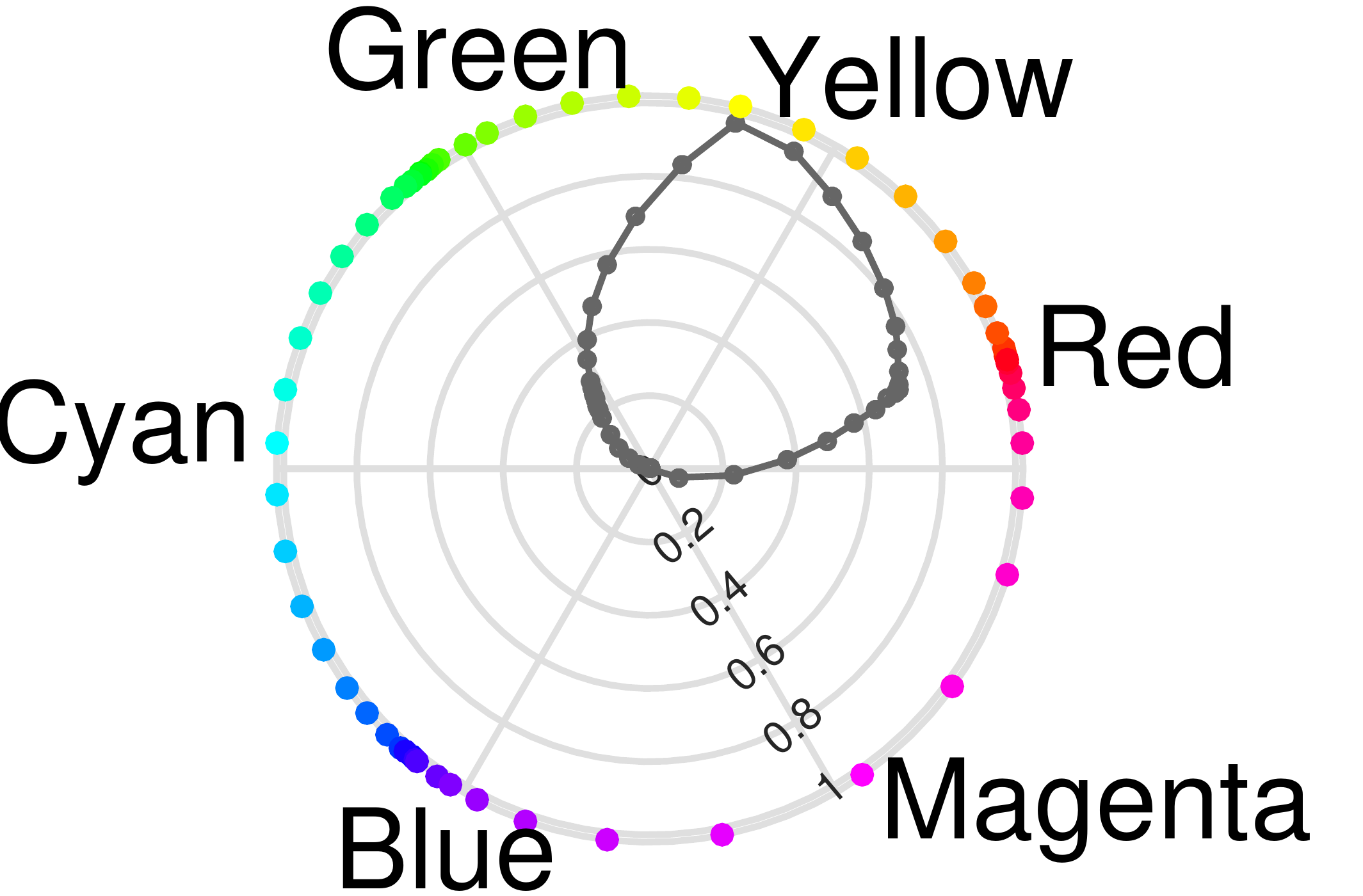}}~
	\subfigure[L-off $\times$ S-on]{\includegraphics[width=0.15\textwidth]{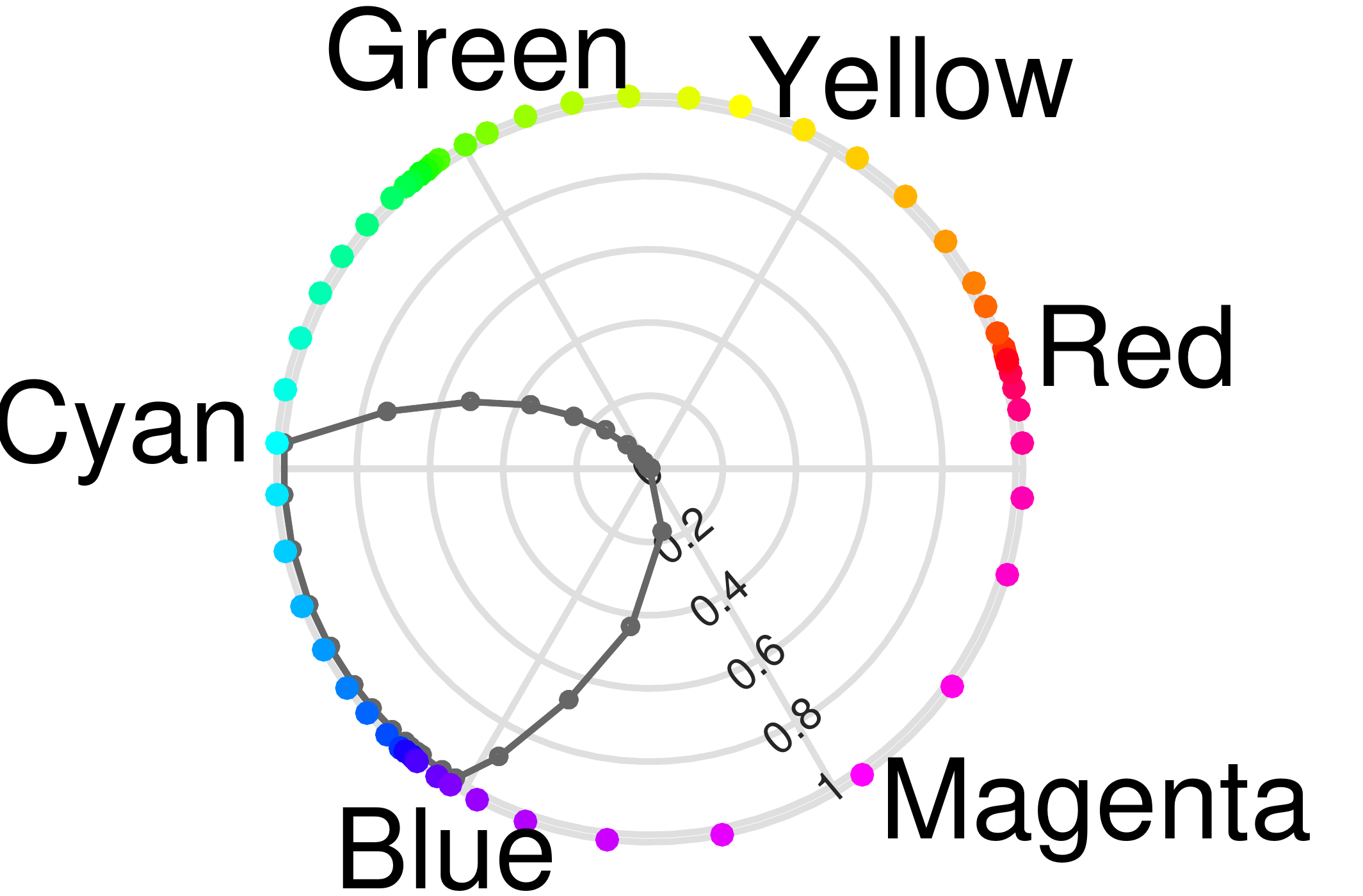}}~
	\subfigure[L-off $\times$ S-off]{\includegraphics[width=0.15\textwidth]{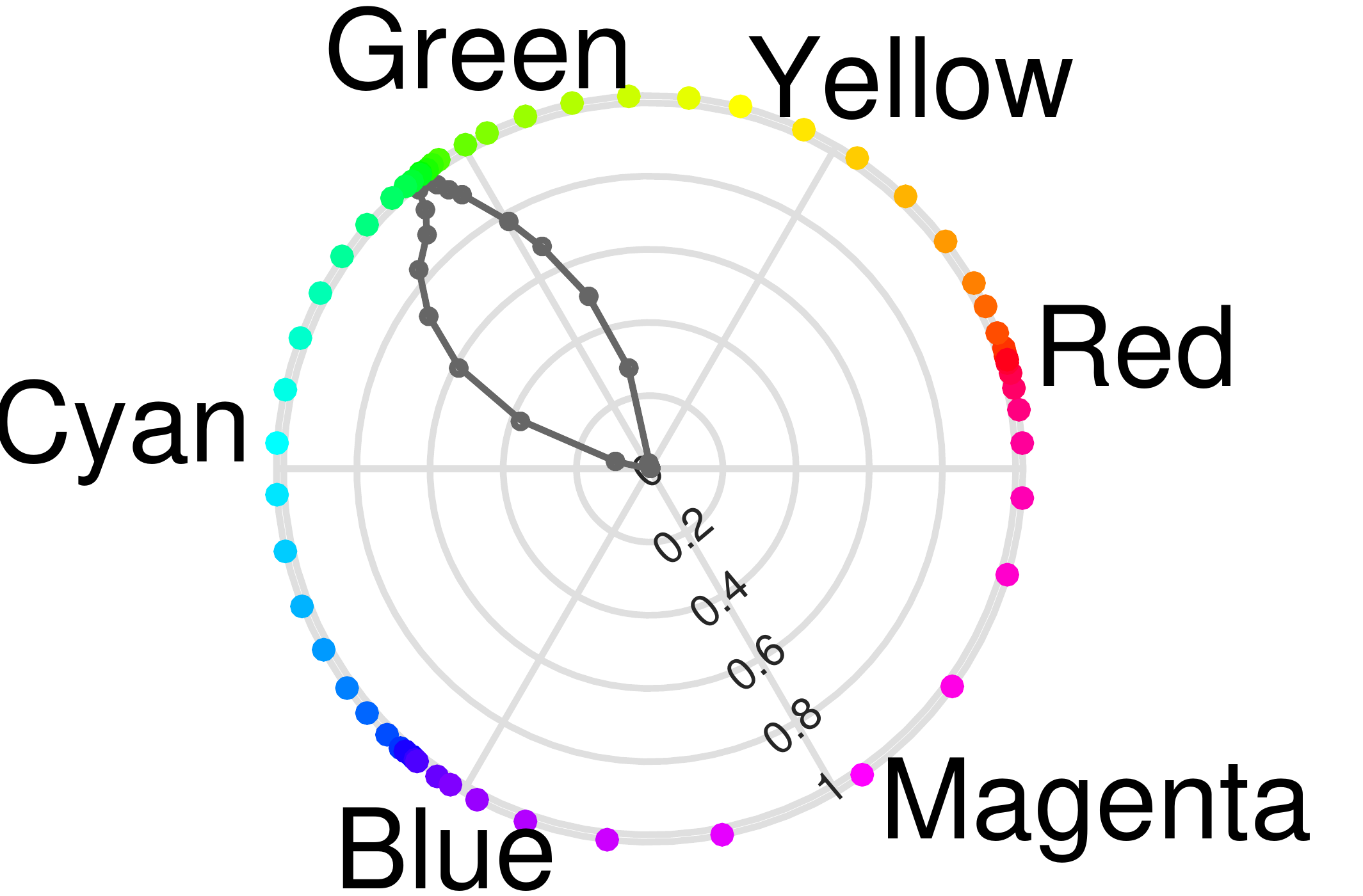}}~
	\subfigure[M-on $\times$ S-on]{\includegraphics[width=0.15\textwidth]{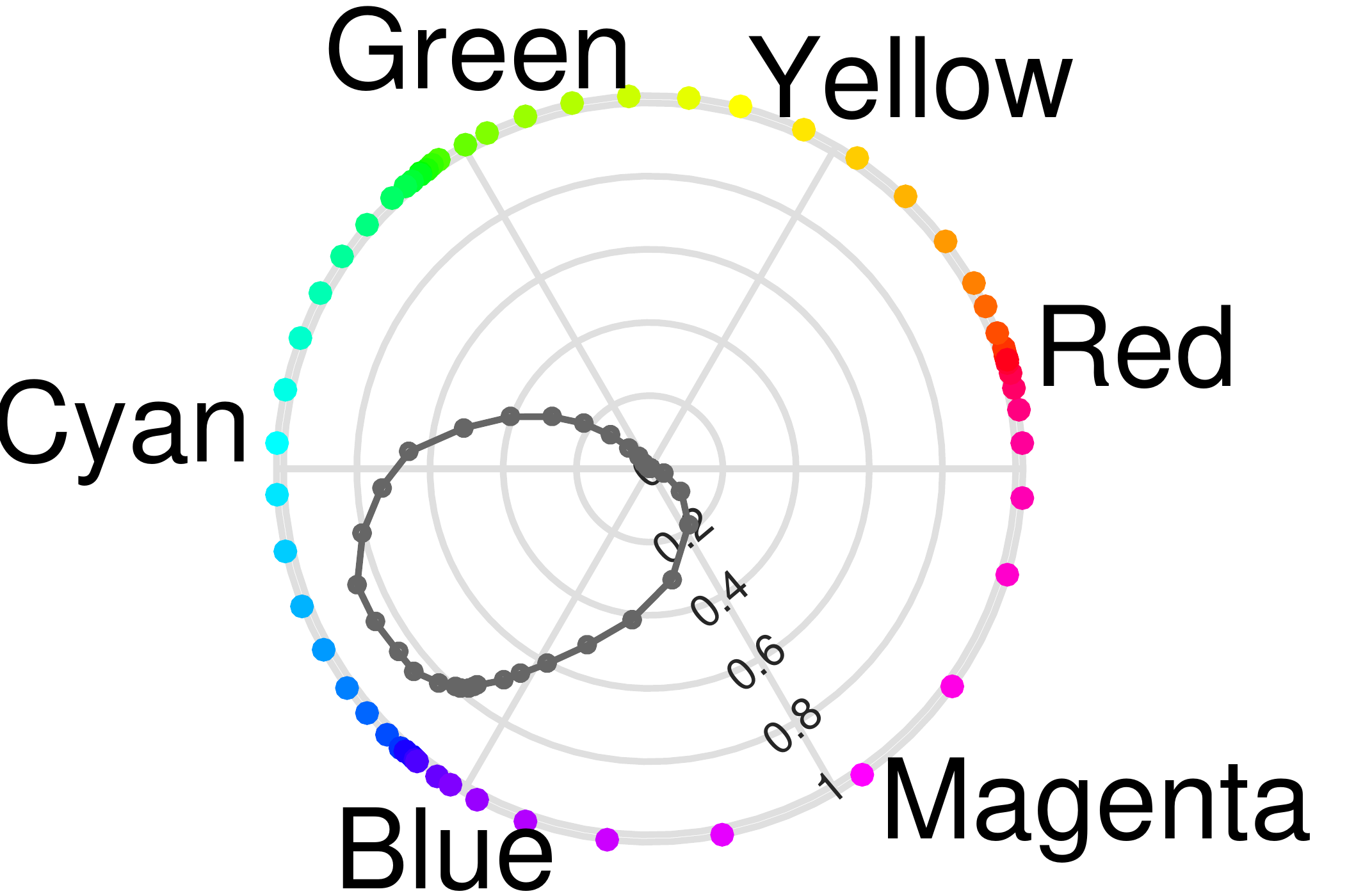}}~
	\subfigure[M-on $\times$ S-off]{\includegraphics[width=0.15\textwidth]{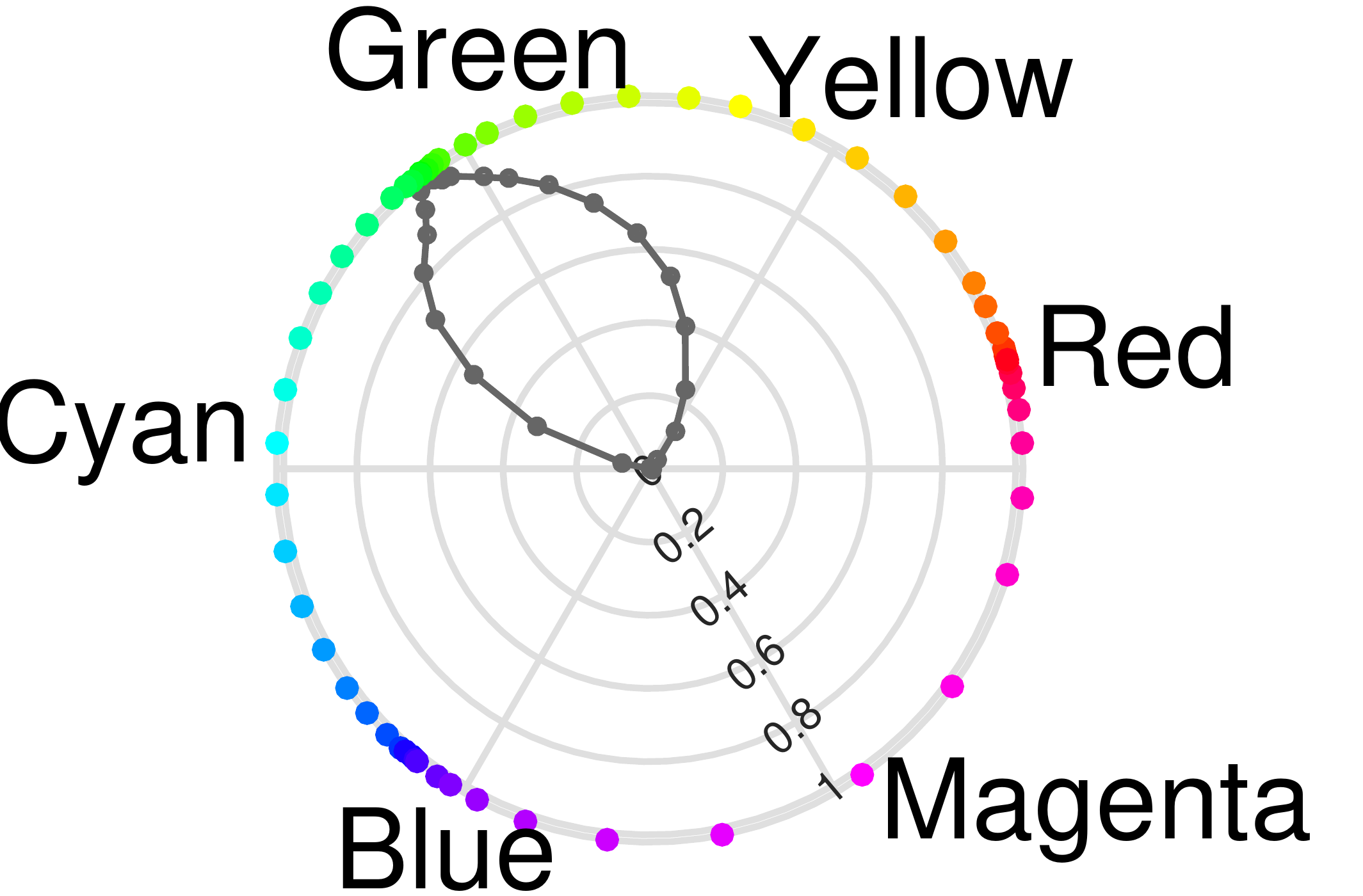}}\\
	\shiftleft{168pt}{Multiplicative V2}
	\subfigure[M-off $\times$ S-on]{\includegraphics[width=0.15\textwidth]{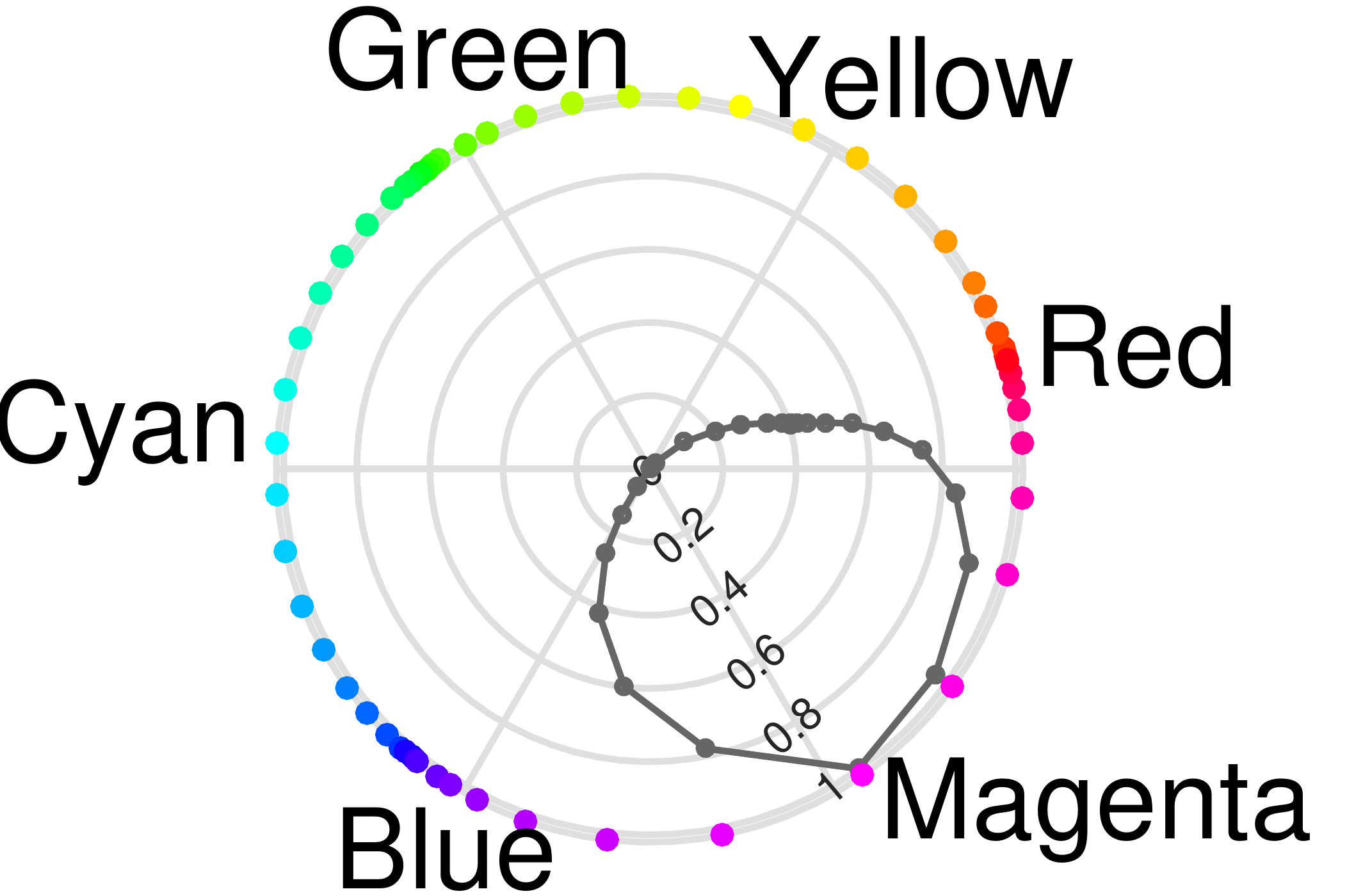}}~
	\subfigure[M-off $\times$ S-off]{\includegraphics[width=0.15\textwidth]{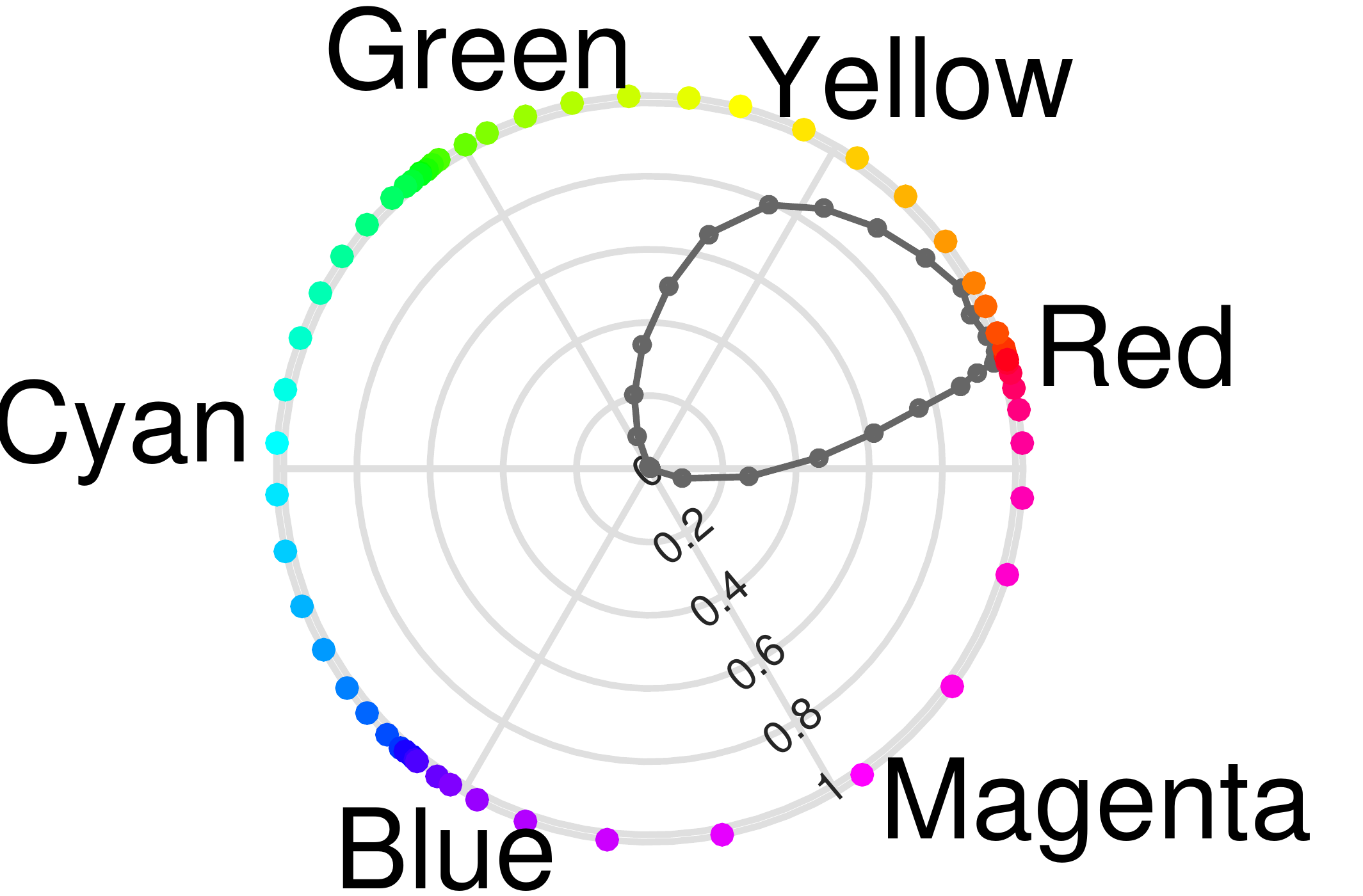}}\\
	\hrulefill
	\vspace{0.2cm}
	\\
	\shiftleft{160pt}{Examples of }
	\shiftleft{152pt}{single-opponent V1}
	\shiftleft{144pt}{and V2 tunings}
	\shiftleft{136pt}{plotted in the}
	\shiftleft{128pt}{$[-1, 1]$ range}
	\subfigure[V1 L-on]{\includegraphics[width=0.15\textwidth]{V1_L_on_activations_MB.pdf}\label{subfig:V1_L_on_neg}}~
	\subfigure[V2 L-on]{\includegraphics[width=0.15\textwidth]{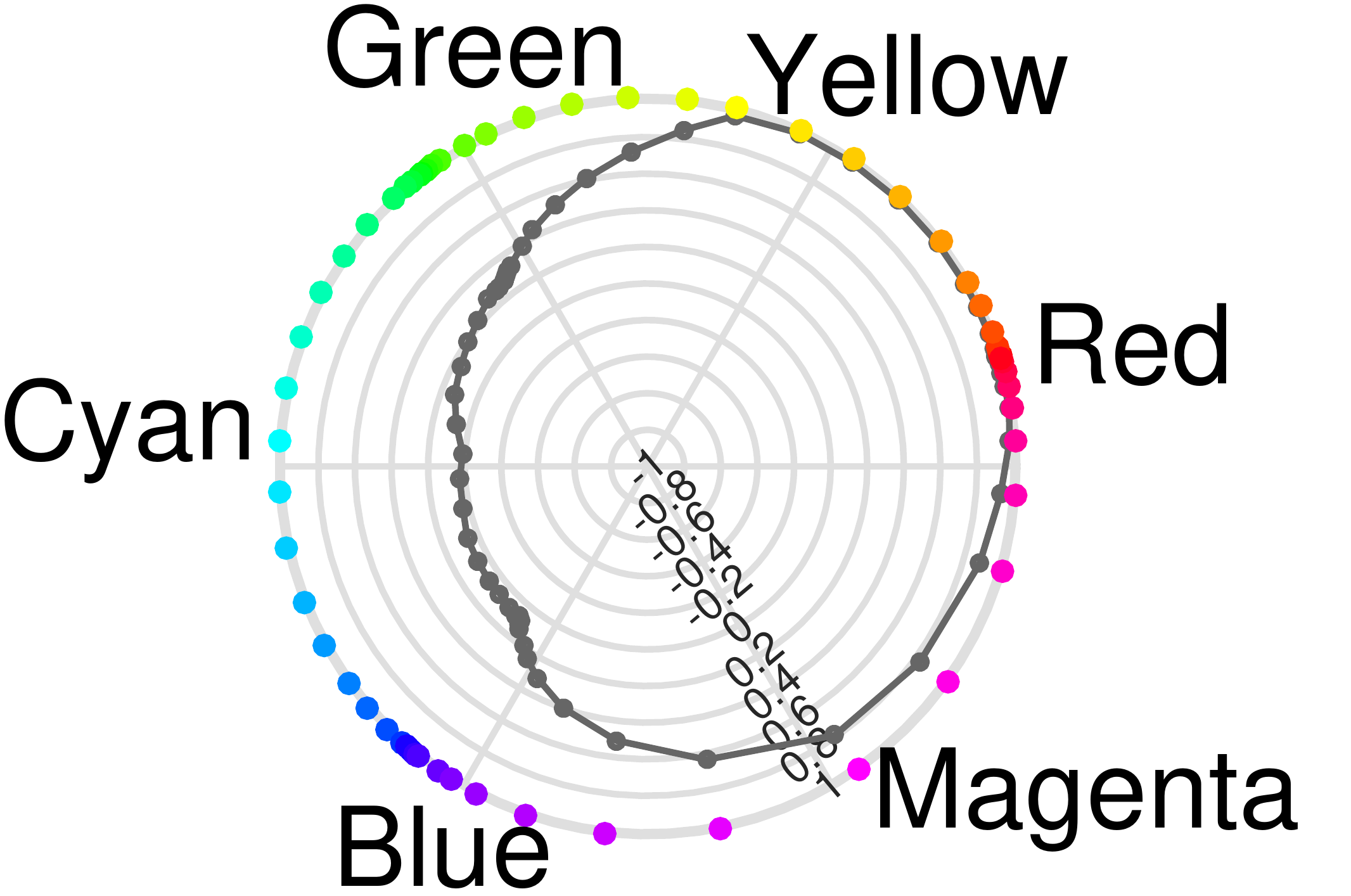}\label{subfig:V2_L_on_neg}}~
	\caption{Responses of neurons in model layers LGN, V1, and V2 to 60 hues sampled from the hue dimension in the HSL space. Each sampled hue is mapped to its corresponding hue anlge in the MB space and is shown by a colored dot corresponding to the sampled hue on the circumference of a unit circle in the MB space. In each plot, the circular dimension represents the hue angle in the MB space. The level of responses is shown in the radial dimension in these plots. In each row, the model layer the neurons belong to is specified on the left edge of the row. The neuron type is mentioned below each plot. Tunings in \subref{subfig:V1_L_on_neg} and \subref{subfig:V2_L_on_neg} show example tunings for single-opponent V1 and V2 cells plotted in the $[-1, 1]$ range. The tunings of these cells, when plotted in the same range, look almost identical.}
	\label{fig:hue_resp_V2}
\end{figure*}

In order to test the effectiveness of our approach for modeling local hues, we examined the tuning of each hue-selective neuron in the individual layers of our network. For this purpose, we sampled the hue dimension of the HSL space. We keep saturation and lightness values constant and set to 1 and 0.5 respectively, following \cite{li2014Map}. Our sampling consists of 60 different hues in the range of $[0, 360)$ degrees, separated by 6 degrees. When these HSL hue angles are mapped to a unit circle in the MB space, they are not uniformly spaced on the circle and are rotated. For example, the red hue in the HSL space at 0 deg corresponds to the hue at about 18 deg in the MB space. The mapping of the 60 sampled hues on a unit circle in the MB diagram are shown on the unit circles in Figure \ref{fig:hue_resp_V2}. The color of each dot corresponds to the hue it represents on the unit circle. Note that the positive vertical direction in the tuning plots corresponds to lime hues, following the plots from Conway and Tsao \cite{Conway_Tsao2009} , their Figure 1. 

\begin{figure*}[p]
	\centering
	\subfigure[magenta]{\includegraphics[width=0.15\textwidth]{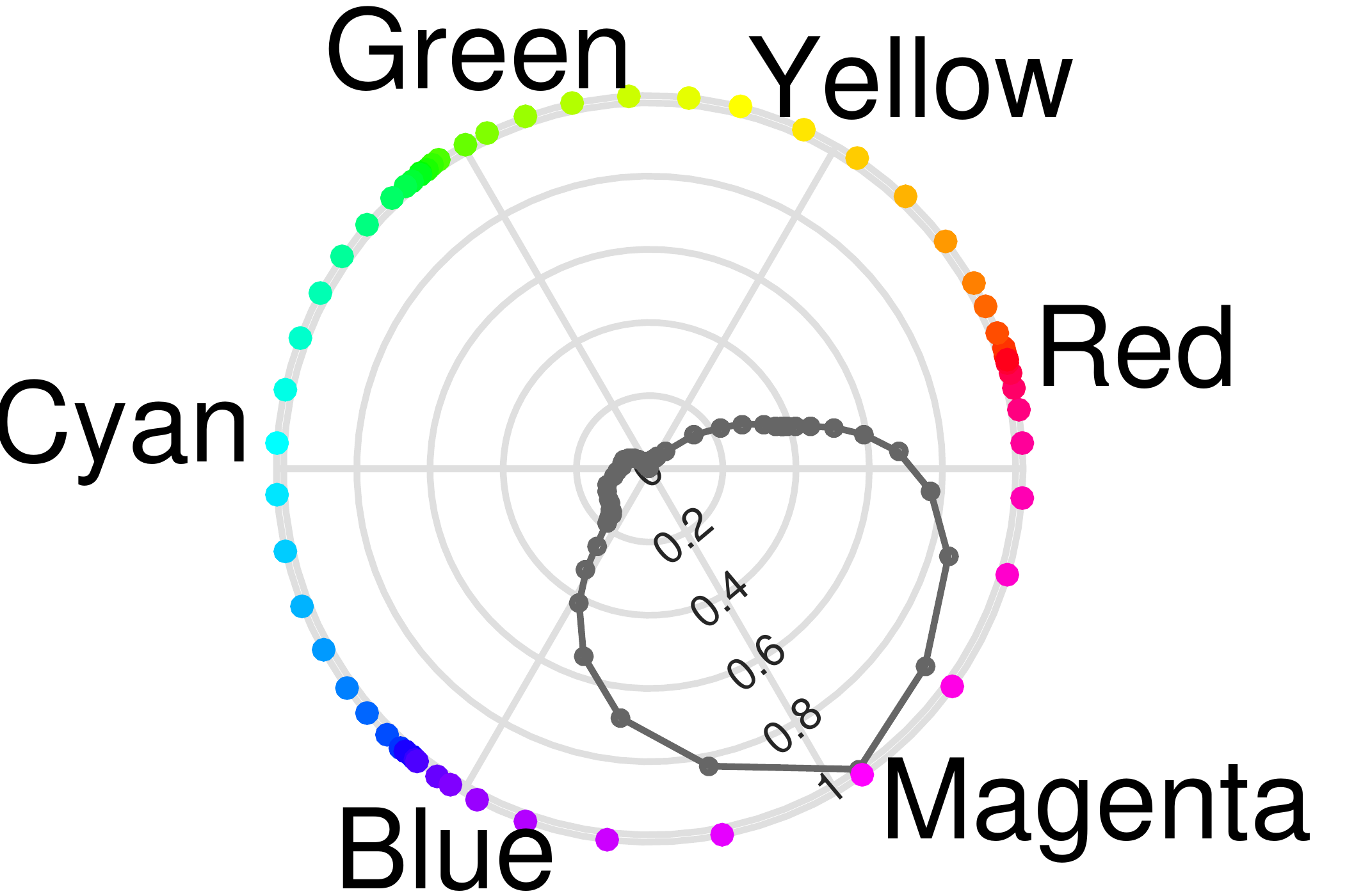}}~
	\subfigure[red]{\includegraphics[width=0.15\textwidth]{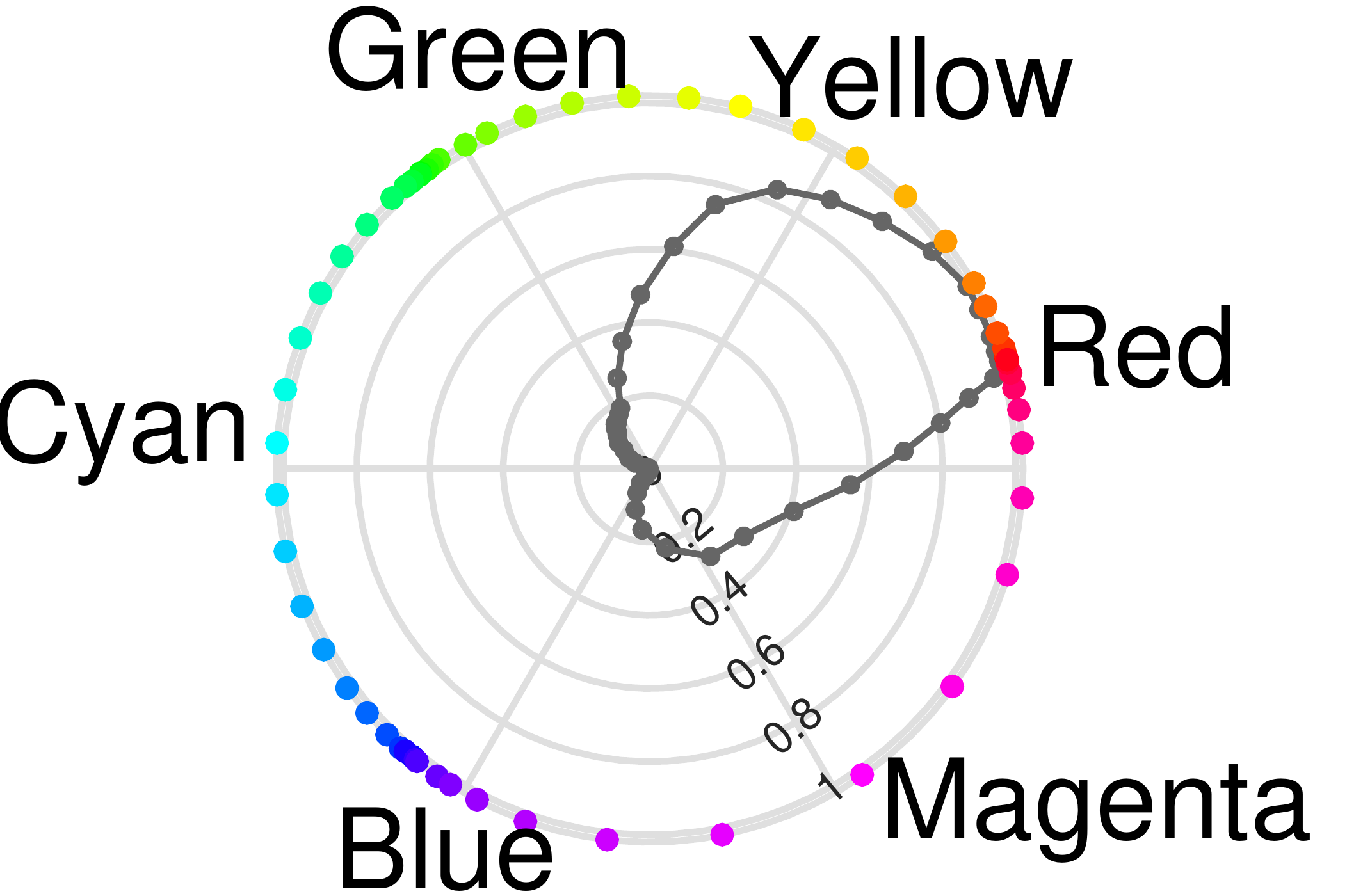}}~
	\subfigure[yellow]{\includegraphics[width=0.15\textwidth]{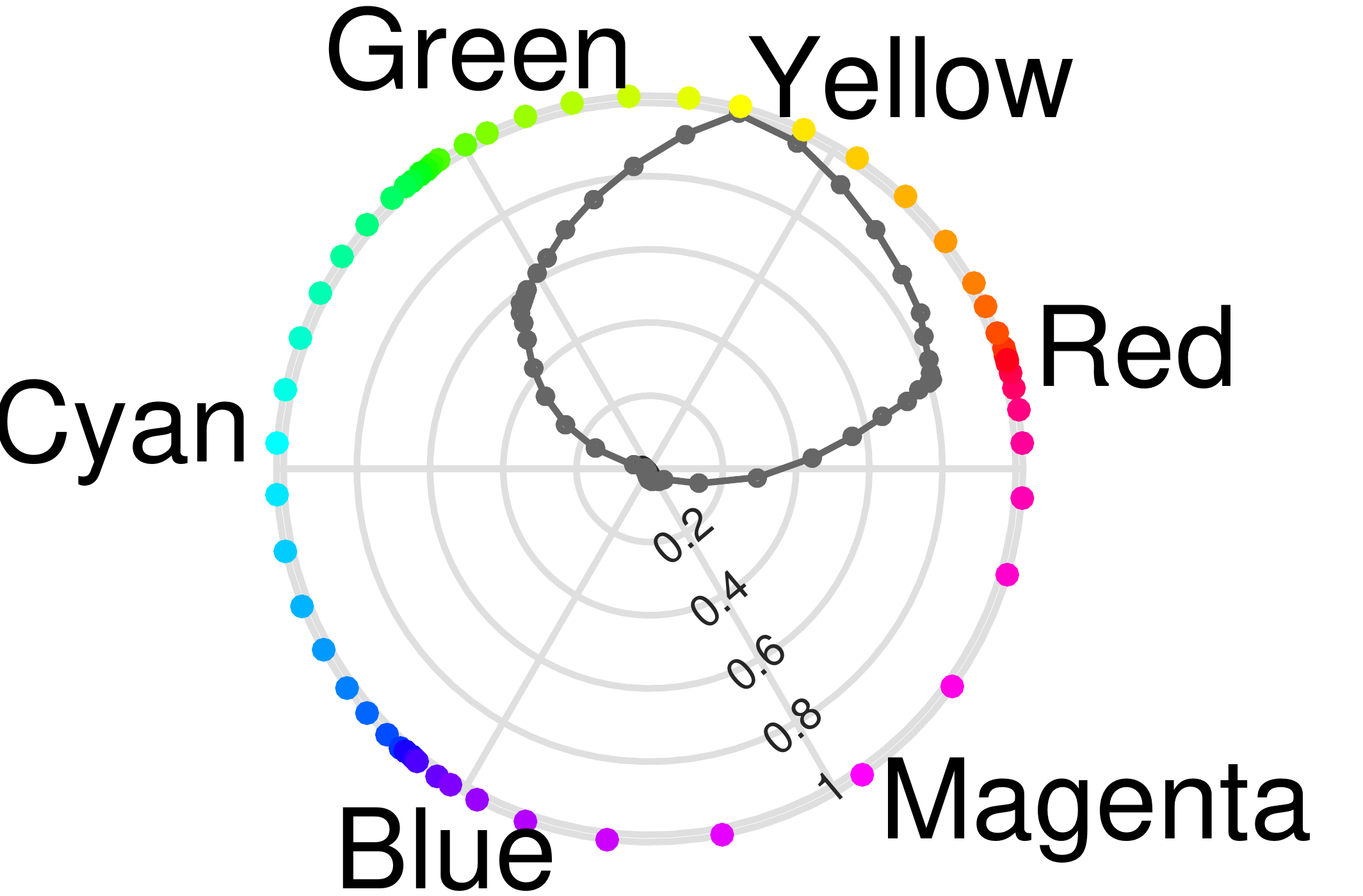}}~
	\subfigure[green]{\includegraphics[width=0.15\textwidth]{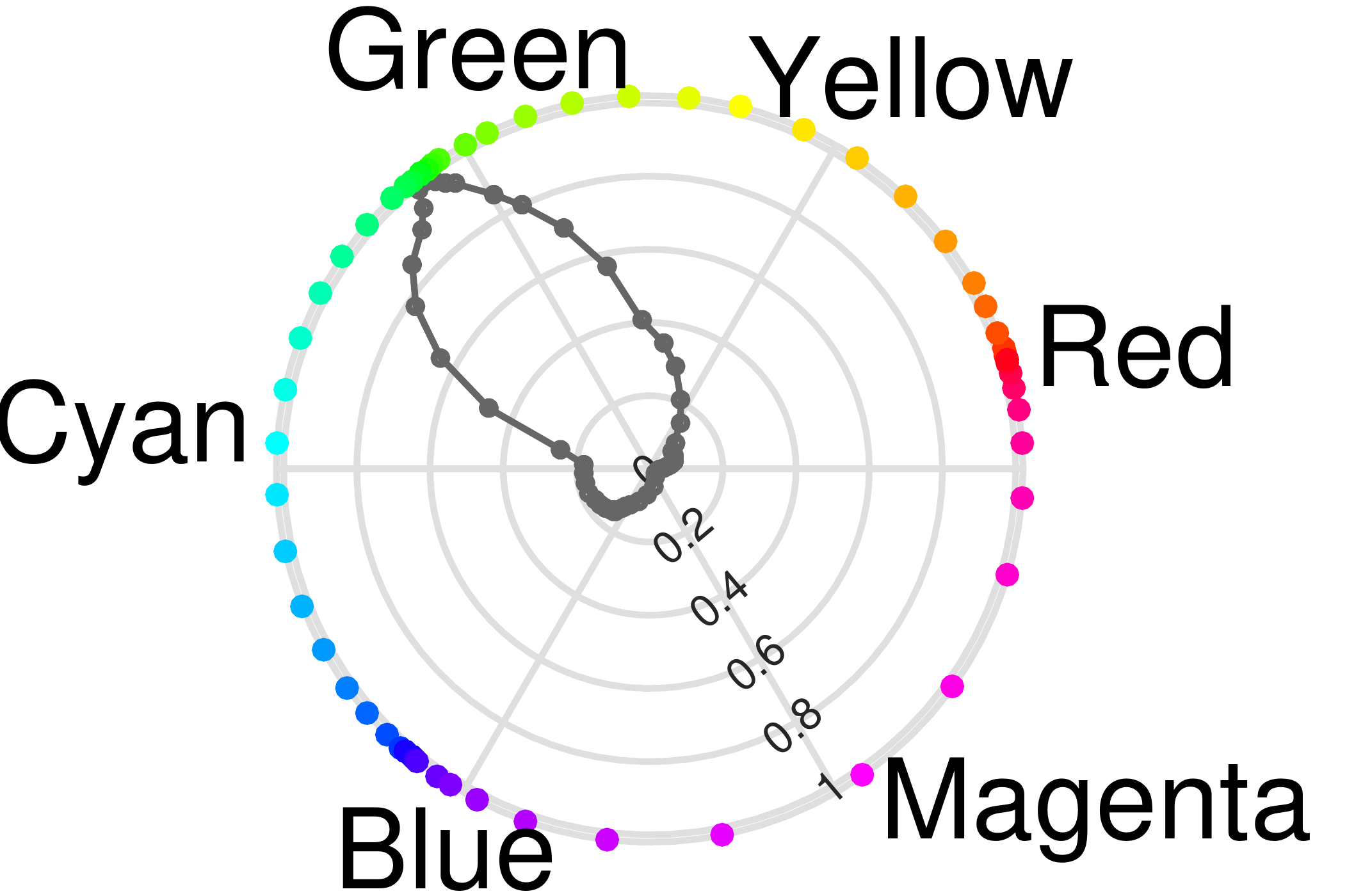}}~
	\subfigure[cyan]{\includegraphics[width=0.15\textwidth]{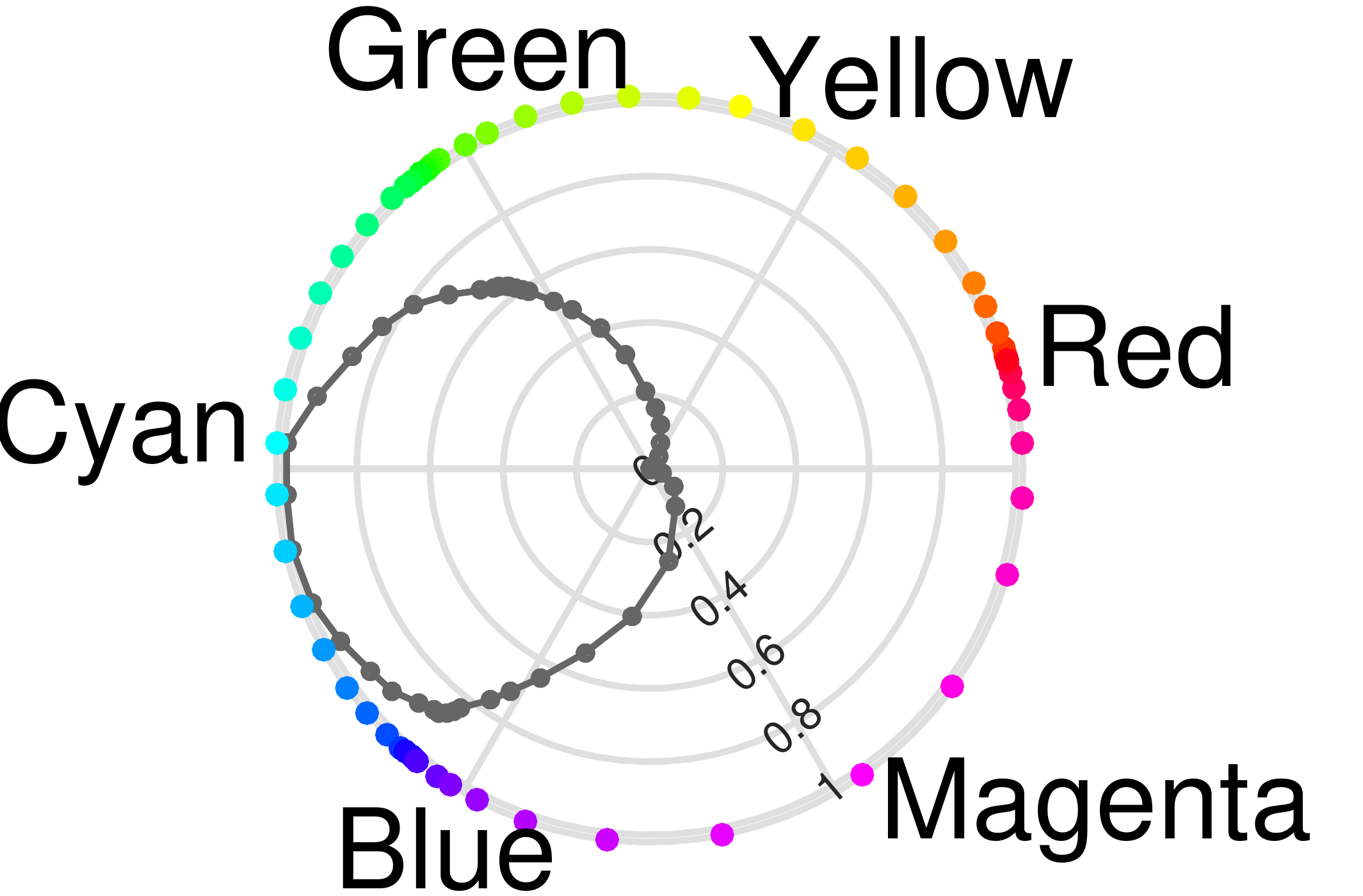}}~
	\subfigure[blue]{\includegraphics[width=0.15\textwidth]{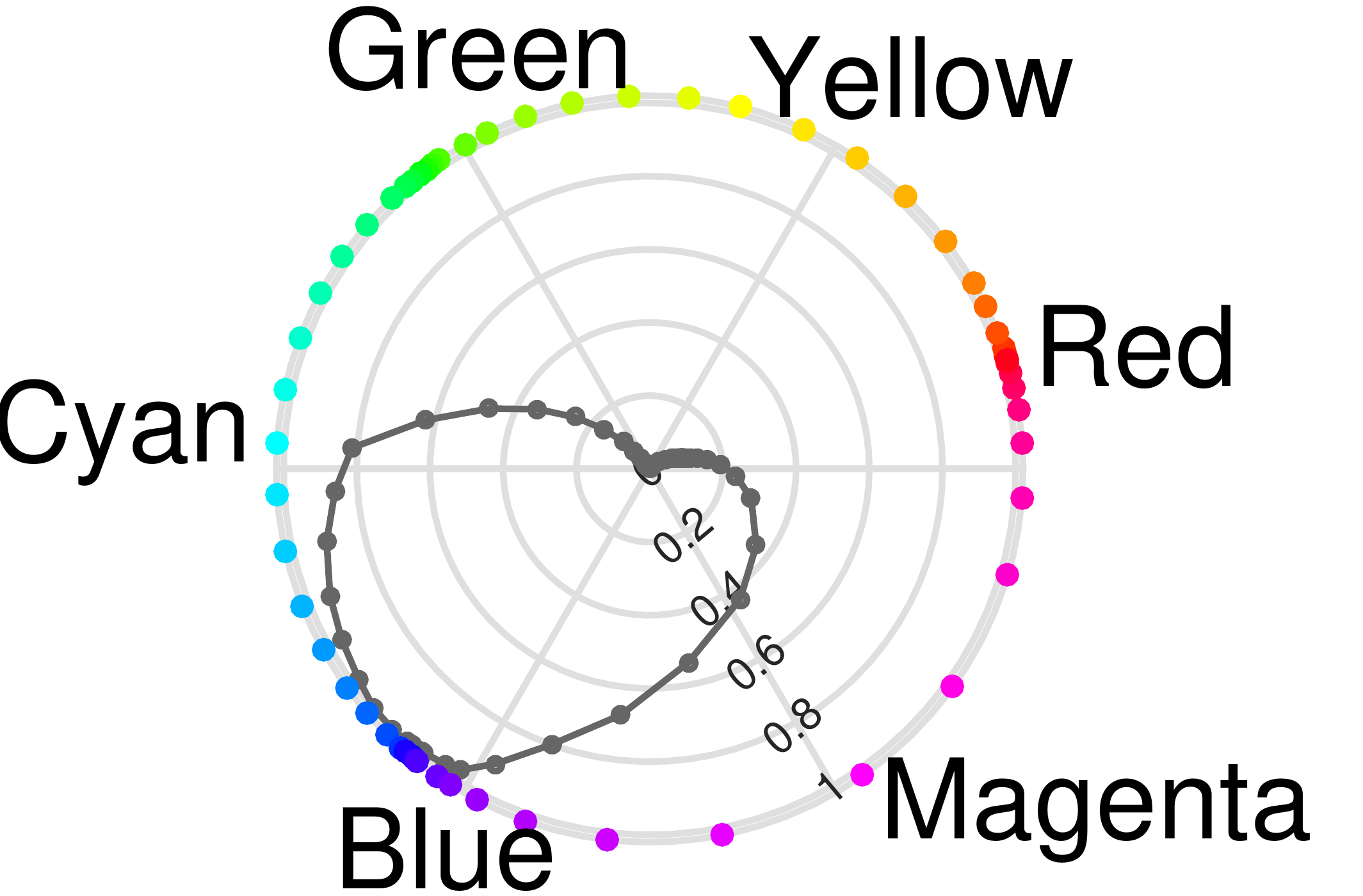}}\\
	\caption{Model V4 neuron responses to hues sampled from the hue dimension in the HSL space. The sampled hues are 6 degrees apart. In each plot, the angular dimension shows the hue angles in the MacLeod and Boyton diagram, and the radial dimension represents the response level of the neuron. The dots on the circumference of each circle are representative of hues at sampled angles. The neuron type is specified below each plot.}
	\label{fig:hue_resp_V4}
\end{figure*}

We present each of these 60 hues to the model and record the activities of model LGN, V1, V2 and V4 neurons. Plots in Figures \ref{fig:hue_resp_V2} and \ref{fig:hue_resp_V4} show model neuron activities to each of the sampled hues. 
In each plot, the circular dimension represents the hue angle in the MB diagram, and the radial dimension represents the response level of the neuron. We found the tunings of our model LGN and V1 neurons look relatively similar with differences due to the nonlinearity of V1 neurons imposed by the rectifier. We plotted the responses of V1 neurons in both negative and positive ranges for comparison purposes with those of LGN, and in V2 and V4, the responses are shown in the positive range. Although it might not be evident from tunings of V1 cells and those of single-opponent V2 neurons in Figure \ref{fig:hue_resp_V2}, due to the plotted range of responses in these figures, we emphasize that the tunings of these cells look identical when plotted in the same range. The average difference of responses between pairs of model V1 and their corresponding single-opponent V2 cells is on the order of $10^{-6}$. An example of similar tunings for these cells is shown in figures \ref{fig:hue_resp_V2} \subref{subfig:V1_L_on_neg} and \ref{fig:hue_resp_V2} \subref{subfig:V2_L_on_neg}

Comparing the tunings of model single-opponent and multiplicative V2 cells give a clear image of narrower tunings in the S-modulated V2 cells. Not only these cells have narrower tunings, but they generally peak close to intermediary directions. See, for example, the tuning of V2 L-off $\times$ S-off cell with a narrow tuning and a peak close to the unique green hue angle.

In our model layer V4, we implemented six different neuron types according to distinct red, yellow, green, cyan, blue and magenta in the HSL space. The chosen hues are 60 deg apart on the hue circle of HSL and the weights from model V2 cells to model V4 neurons were determined according to the distance between mean peak activations of model V2 neurons to the desired hue in a model V4 cell. Tunings of our model V4 neurons, depicted in Figure~\ref{fig:hue_resp_V4}, show a clear peak for each cell close to its desired selectivity, with narrower tunings compared to single-opponent V2 cells. 

\subsection*{Tuning bandwidths}
\begin{figure*}[tph!]
	\centering
	\subfigure[]{\includegraphics[width=0.45\textwidth]{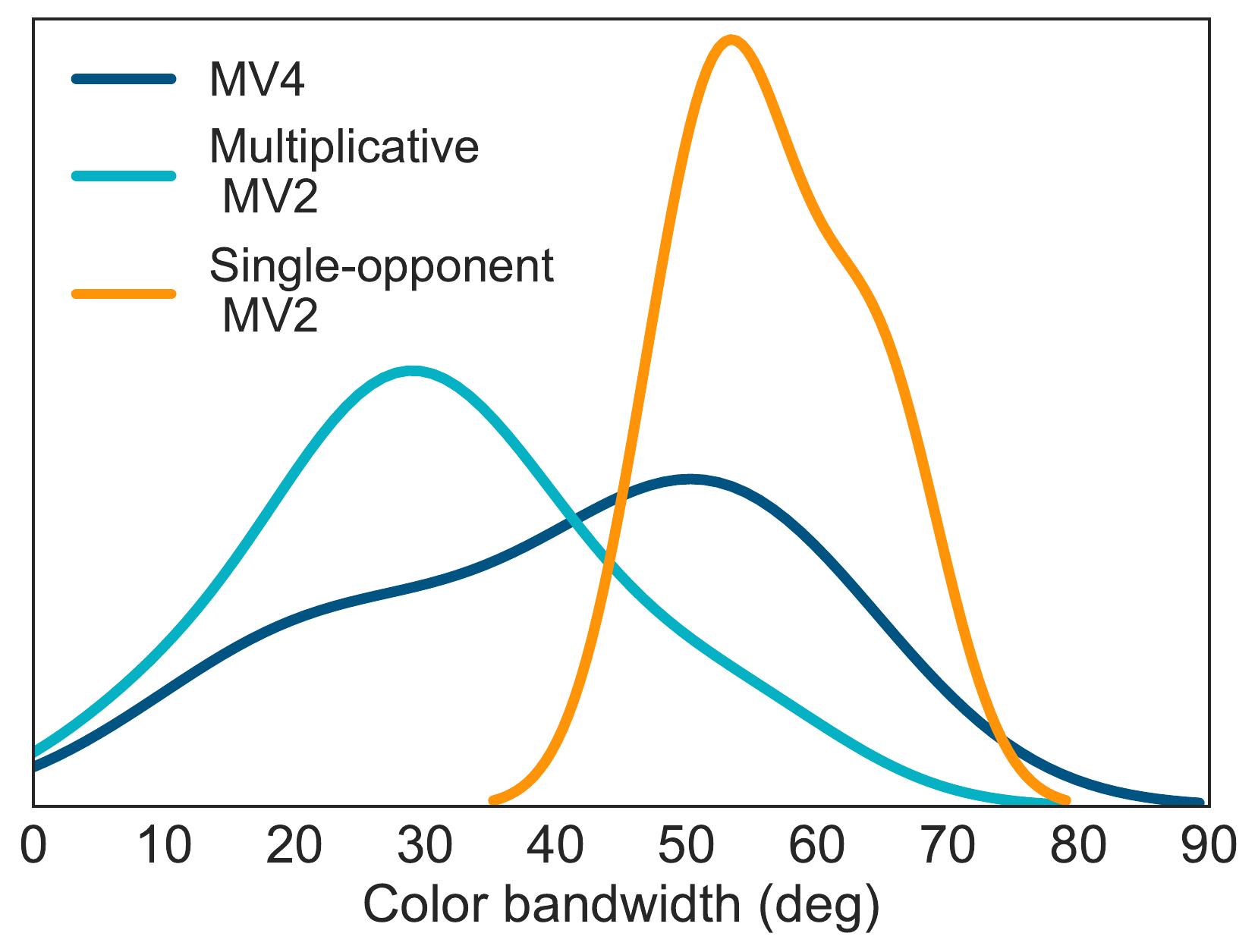}\label{subfig:bandwidth}}
	\subfigure[]{\includegraphics[width=0.54\textwidth]{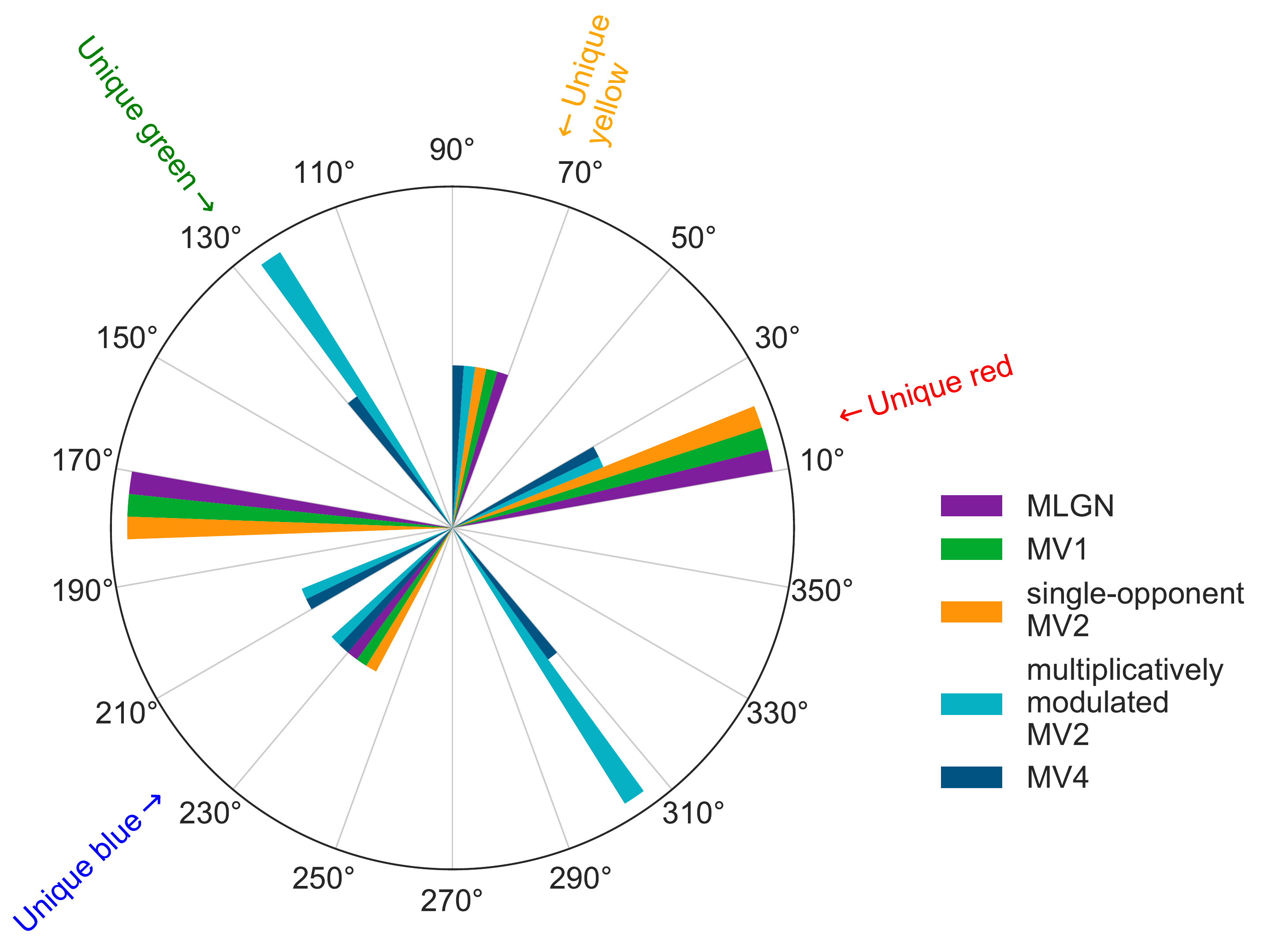}\label{subfig:mean_peak}}
	\caption{\subref{subfig:bandwidth} Distributions of tuning bandwidths for neurons in model layers V2 and V4 called MV2 and MV4 respectively. Note that neurons in model layers LGN and V1 have tunings similar to those of single-opponent V2 and we refrain from including those cells in this figure for simplicity. The distribution of multiplicative V2 neurons is clearly separate and shifted toward smaller bandwidths compared to single-opponent V2 cells. This confirms our observation that multiplicative modulations result in narrower tunings. In V4, cells with very small bandwidths, as small as 16 deg, were observed, while neurons with larger bandwidths were also found in this layer. Note that our model V4 cells are linear combinations of both multiplicative and single-opponent cells and depending on the weights from V2 cells, the bandwidth of these neurons might vary between small to mildly large. \subref{subfig:mean_peak} A polar histogram of mean selectivity peaks for our model neurons. Our model LGN, V1 and single-opponent V2 neurons cluster around cone-opponent axes directions while our model multiplicative V2 and V4 cells have mean peaks both close to cone-opponent and off-opponent hue directions. Note specifically that all model layers have neurons in polar bins containing unique red and unique yellow hue angles. Bins including unique green and unique blue angles are limited to multiplicative V2 and V4 types.}
	\label{fig:bandwidth}
\end{figure*}

In order to obtain a quantitative evaluation of the above observation with regards to narrower tunings due to multiplicative modulations, we computed the bandwidth of all  neurons in model layers V2 and V4, following \cite{Kiper1997V2}. In case of peaks at more than one hue, we take the mean peak hue as the peak response representative for computation of bandwidth and later for peak tunings. Note that the goal of this analysis is to verify whether multiplicative modulations result in neurons with smaller bandwidths, \ie narrower tunings. In this work, we did not hypothesize a model as Kiper \etal \cite{Kiper1997V2} who computed the bandwidth threshold analytically for linear and nonlinear tunings, nor did we have a population of neurons to report the percentage of linear/nonlinear cells. Instead, we simply computed the bandwidth for each cell type in model layers V2 and V4 with respect to its responses at discrete sampled hues and plotted the distribution of tuning bandwidths. Specifically, we computed the bandwidth of 6 single-opponent V2 cells, 8  multiplicatively modulated V2 neurons, and 6 hue-selective V4 cells tunings that were shown in Figures \ref{fig:hue_resp_V2} and \ref{fig:hue_resp_V4}. The bandwidth distributions are depicted in Figure \ref{fig:bandwidth}. 
Our model single-opponent and multiplicative V2 cells have clearly separate distributions, similar to observations of Kiper \etal \cite{Kiper1997V2} for linear and nonlinear neurons. Single-opponent V2 neurons are mainly clustered around large bandwidths, between 48 and 66 deg with mean at 56.53 deg. Bandwidths of multiplicative V2 cells vary between 9 and 55 deg with an apparent density toward bandwidths smaller than 40 deg and mean at 30.84 deg. These results confirm our previous observation that multiplicative modulations lead to narrower tunings. Here, we skipped plotting distributions for LGN and V1 neurons as they are essentially similar to single-opponent V2 cells. In model layer V4, bandwidths vary between 16 and 58 deg with mean at 41.04 deg, a range similar to multiplicatively modulated V2 bandwidth range. In this layer, however, the shift toward larger bandwidths in comparison to multiplicative V2 cells is not surprising as V4 neuron activations are a weighted sum of both multiplicative and single-opponent V2 cells. As a result, depending on the contribution from each type of V2 cell, a V4 neuron might have narrow to mildly wide tunings.

\subsection*{Tuning mean peaks}
Watcher \etal \cite{Wachtler2003} observed that most neurons in V1 peak around non-opponent directions, while Kiper \etal \cite{Kiper1997V2} reported that cells in V2 exhibited no preference to any particular color direction and no obvious bias to unique hues. We tested the mean tuning peaks of our model neurons to examine for cone-opponent vs. intermediate selectivites. Figure \ref{subfig:mean_peak} shows a polar histogram of tuning mean peaks of neurons in all layers of our model, where each sector of the circle represents a bin of the polar histogram. This figure clearly demonstrates that the majority of LGN, V1, and single-opponent V2 cells (5 out of 6 neuron types) peak close to cone-opponent axes. In contrast, our model multiplicative V2 cells and hue-sensitive V4 neurons peak at both cone-opponent and intermediate hues, as reported in \cite{Kiper1997V2} and \cite{Bohon2016}. In other words, with an increase in nonlinearity, representation of hues along intermediate directions start to develop.

\subsection*{Unique hue representation}
\begin{figure*}[p]
	\centering
	\subfigure[]{\includegraphics[width=0.4\textwidth]{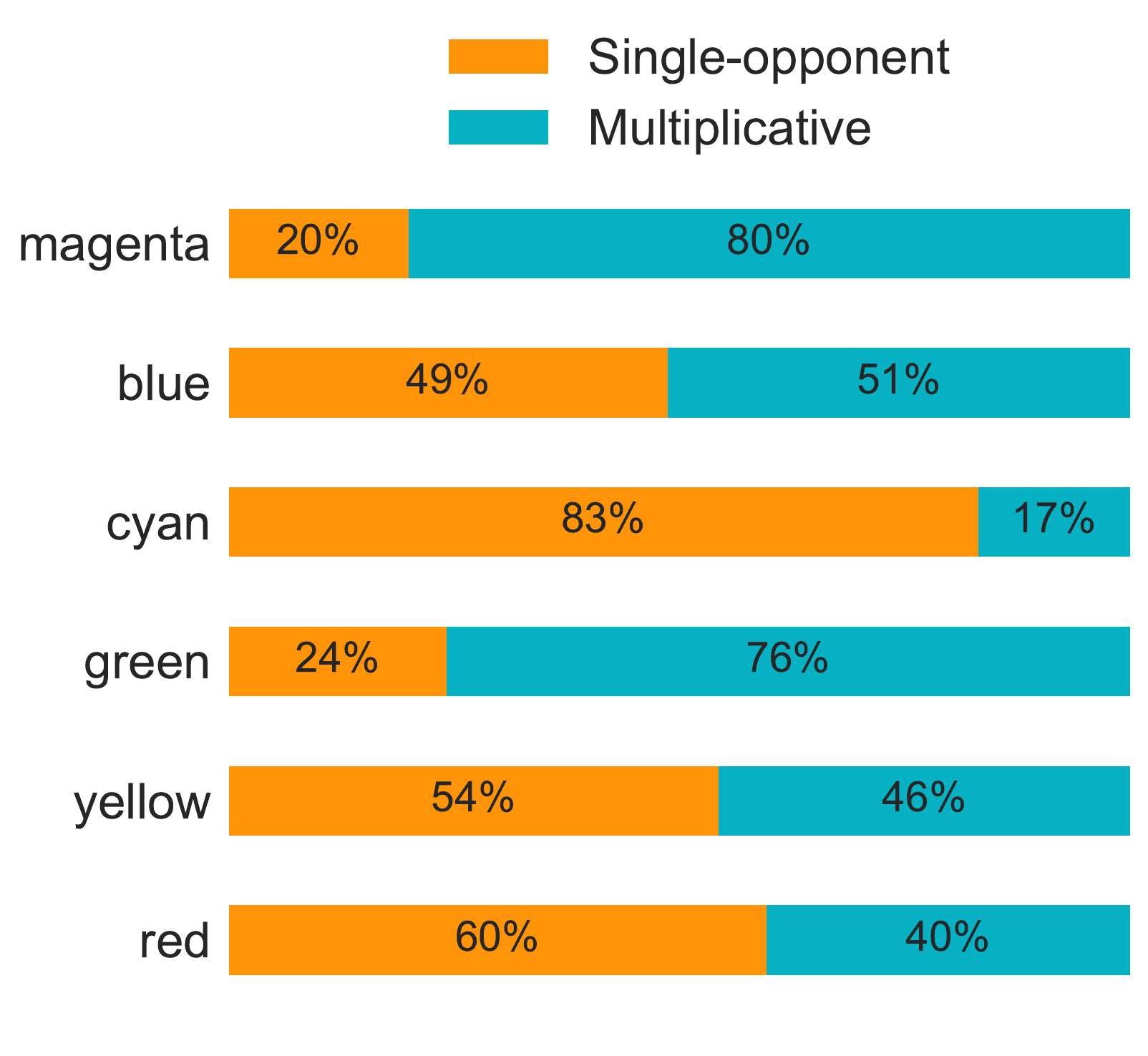}\label{subfig:single_multi_weight}}
	\subfigure[]{\includegraphics[width=0.45\textwidth]{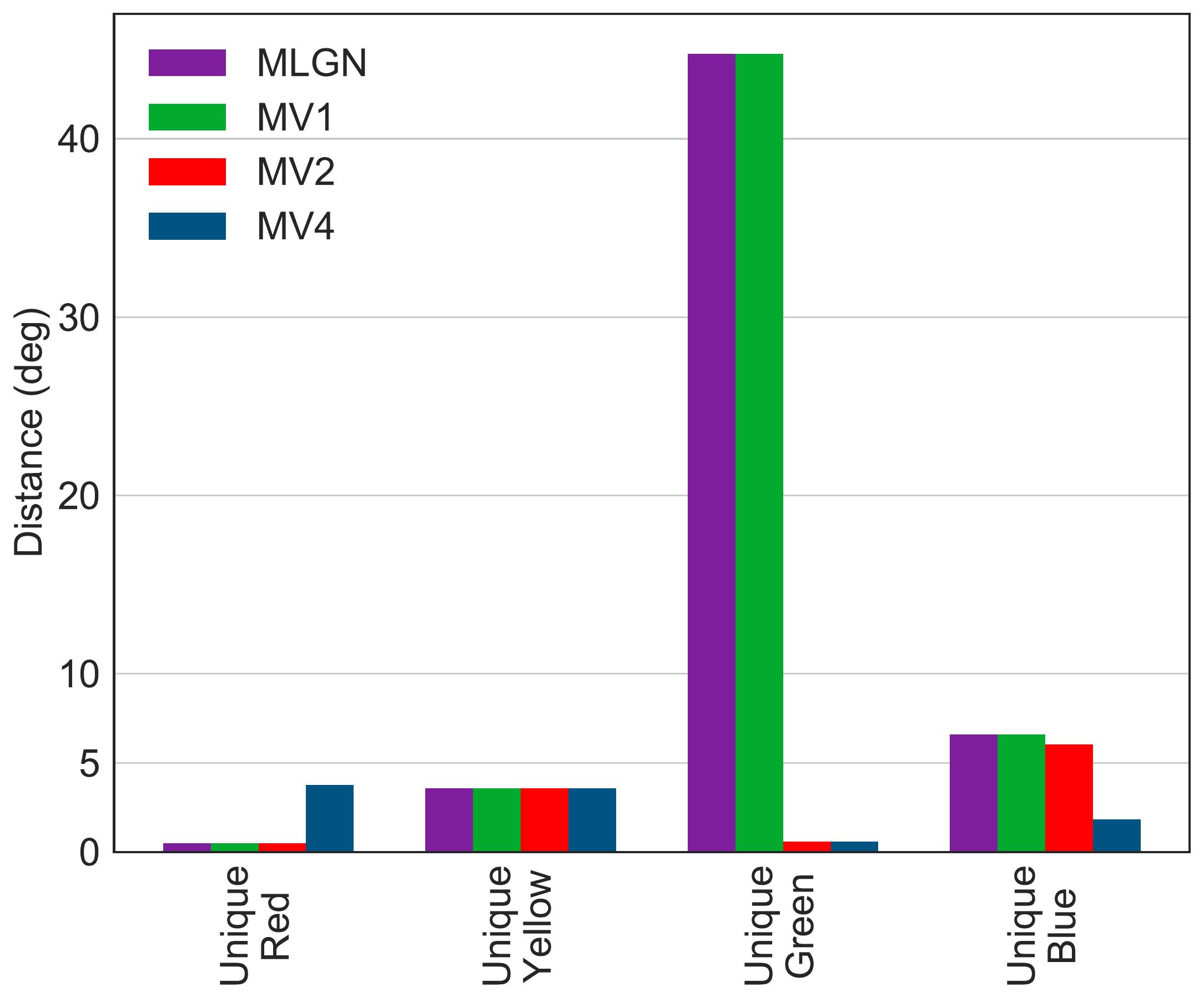}\label{subfig:unique_hue_distances}}\\
	\subfigure[]{\includegraphics[width=0.55\textwidth]{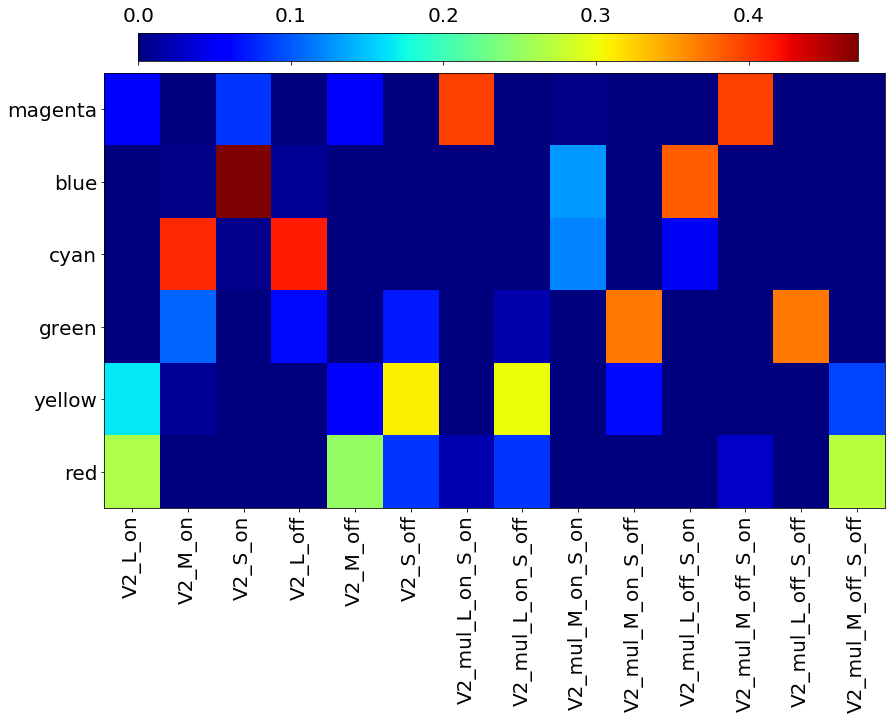}\label{subfig:V2_V4_weights}}
	\caption{\subref{subfig:single_multi_weight} Relative contributions of single-opponent and multiplicative V2 cells to each hue-sensitive V4 neuron. The orange and blue bars show the percentage of contribution from single-opponent and multiplicative V2 cells to each specific V4 cell, respectively. This plot demonstrates that multiplicative V2 neurons have larger input weight to model V4 cells with selectivity in intermediate hue directions and single-opponent cells making up most feedforward input weight to model V4 cells with peaks close to cone-opponent axes. \subref{subfig:unique_hue_distances} Distances of our model neurons, layer-by-layer, to unique hue angles reported by Miyahar \cite{Miyahara2003focal}. For each given model layer and a unique hue, the minimum distance from mean peak hues of all neurons in the layer to the unique hue is reported. As the plot shows, unique hue representation develops in the hierarchy and the distances decrease gradually from LGN to V4. Unique red and yellow representations develop in earlier stages compared to unique green and blue. Note the significant drop in the mean peak distance for V2 and V4 neurons to unique green, achieved by increasing the nonlinearity in those layers. \subref{subfig:V2_V4_weights} The relative weights from model V2 cells to model V4 neurons (best seen in color). In this figure, each row represents a hue-sensitive V4 cell, and each column shows a V2 cell. Multiplicative V2 cells are indicated as V2\_mul\_x\_y, where x and y signify the type of V1 neurons sending feedforward signal to the multiplicative V2 cell. The weights are normalized to sum to 1.0, with dark red indicating a large weight and dark blue for weights close to zero. This figure illustrates the contributions of individual model V2 cells in the encoding of six distinct hues in our model V4 layer. For example, the weights for red confirm that model V2 L-on, M-off, and M-off $\times$ S-off cells, \ie those with large L cone inputs, have the largest contributions. Determining weights from model V2 to V4 layers in our network is described in detail in the Methods section.}
\end{figure*}

In Figure \ref{subfig:mean_peak}, we observed that each V4 bar in the polar histogram was paired with a multiplicative V2 bar. We wondered about the contribution of multiplicative V2 cells to the responses of V4 neurons. Therefore, we compared the sum of single-opponent V2 cell weights against that of multiplicative V2 neurons, depicted in Figure \ref{subfig:single_multi_weight}. Interestingly, V4 cells selective to magenta, blue, and green hues, which are off the cone-opponent directions, have significant contributions from multiplicative V2 cells. In green and magenta, in particular, multiplicative cells make up more than $75\%$ of V2-V4 weights. The rest of hues, which are close to the cone-opponent directions in the MB space, receive a relatively large feedforward input from the single-opponent V2 cells. In short, multiplicative cells play a significant role in the representation of hues in intermediate directions, while single-opponent cells have more substantial contributions to hues along cone-opponent axes.

Next, we asked the question of how well neurons in each of our model layers represent unique hues? To answer this, we computed the distance of mean peak angles for our model neurons to the unique hue angles reported by Miyahara \etal \cite{Miyahara2003focal}. For each unique hue, in each layer of our model, we report the distance of a model neuron with a peak closest to the distinct unique hue angle. That is, the minimum distance of mean peak angle among all neuron types in each layer to a given unique hue angle. The distances, layer by layer, are shown in Figure \ref{subfig:unique_hue_distances}. 
In this figure, the distances of our model neurons to unique red and yellow are relatively small for all the four model layers, at less than 5 deg. This could be due to the fact that unique red and yellow reported in \cite{Miyahara2003focal} are close to cone-opponent axes in the MB space, in agreement with findings of \cite{Webster2000variations} for unique red hue. For unique green and blue, the V2 and V4 distances are far smaller than those of earlier layers. Specifically, there is a 44 deg drop in V2 and V4 distances to unique green compared to that of V1. Also, the distance is even smaller in V4 than V2 to the unique blue hue. As a summary, we observed a gradual development in the exhibition of selectivity to unique green and blue hues, while selectivity to unique red and yellow was observed in early as well as higher layers. Moreover, these results suggest that V4 cells with peak selectivities at less than 5 deg distance to unique hue angles, and consequently, neurons in higher layers, have the capacity to encode unique hues. 

\subsection*{Hue Distance correlation}
\begin{figure}[!tp]
	\centering
	\subfigure[]{\includegraphics[width=0.2\textwidth]{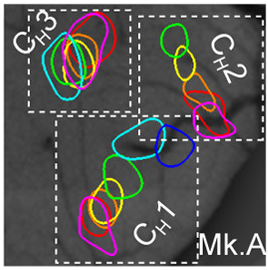}\label{subFig:colorMap_clusters}}\\
	\subfigure[]{\includegraphics[width=0.2\textwidth]{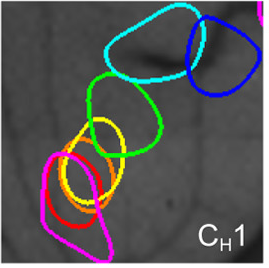}\label{subFig:colorMap_cluster1}} ~
	\subfigure[]{\includegraphics[width=0.2\textwidth]{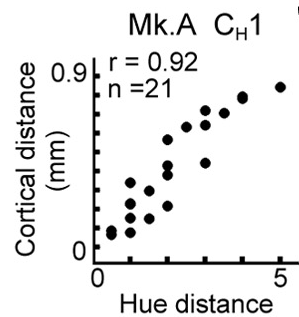}\label{subfig:hue_distance_cluster1}}\\
	\subfigure[]{\includegraphics[width = 0.4\textwidth,clip]{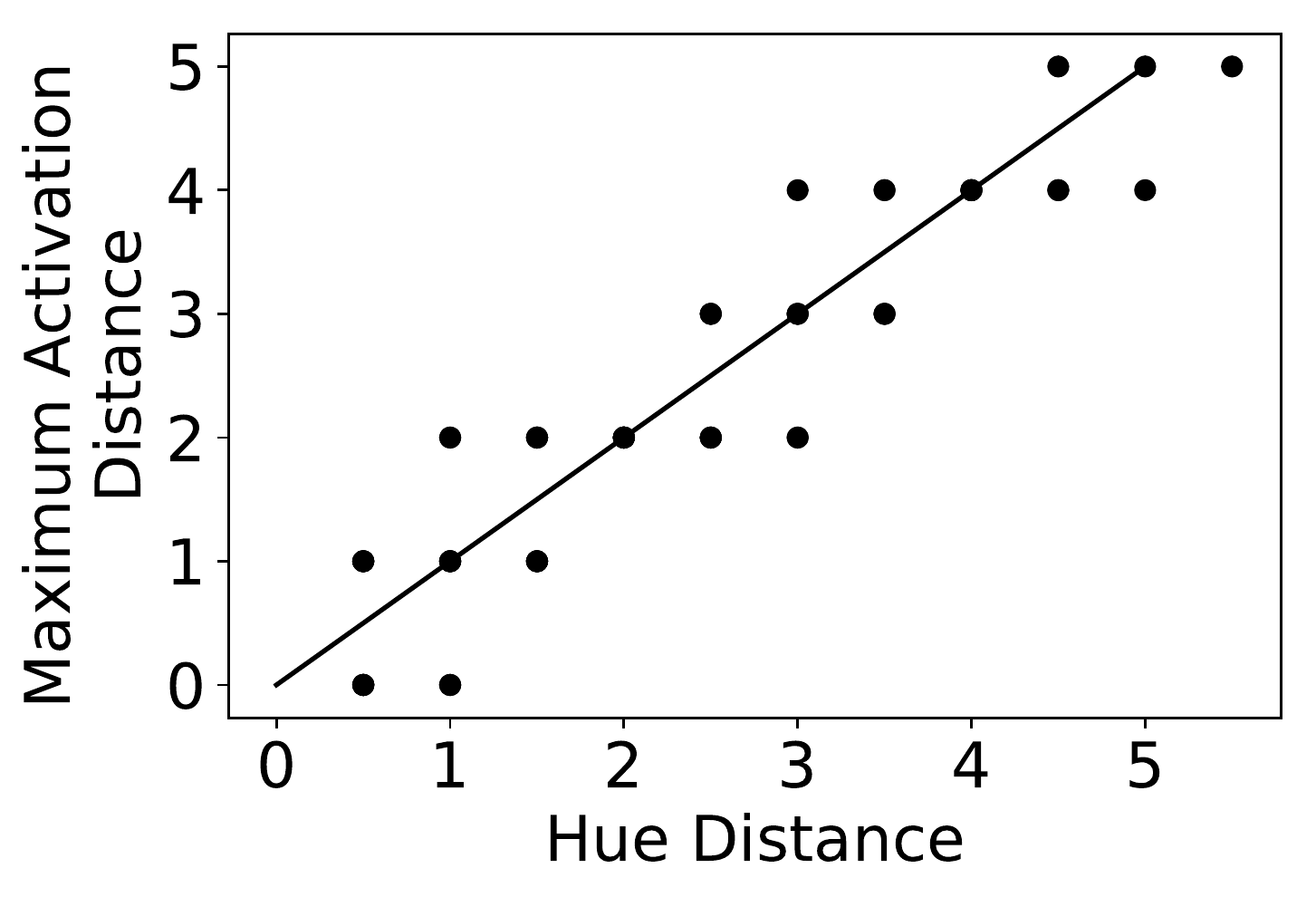}\label{subFig:hue_distance}}
	\caption{\subref{subFig:colorMap_clusters} Color map of V4 neurons in three clusters of patches (adapted from~\protect\cite{li2014Map}). \subref{subFig:colorMap_cluster1} Larger view of a cluster of patches in V4(adapted from~\protect\cite{li2014Map}). \subref{subfig:hue_distance_cluster1} Cortical distance of activated patches in \ref{subFig:colorMap_cluster1} as a function of hue distances (adapted from~\protect\cite{li2014Map}).\subref{subFig:hue_distance} Correlation analysis between the hue distances and model patch distances in each model cluster.}
	\label{fig:colorMap}
\end{figure}

In their study, Li~\etal~\cite{li2014Map} found a correlation between pairs of stimulus hue distances and the cortical distances of maximally activated patches in each cluster. Figure \ref{subFig:colorMap_clusters} illustrates an example of an identified map of three clusters of patches in V4 reported in \cite{li2014Map}, with one cluster shown in a larger view in Figure \ref{subFig:colorMap_cluster1}. Figure \ref{subfig:hue_distance_cluster1} depicts the cortical distances of activated patches for pairs of hues as a function of the stimulus hue distances. 
For this analysis, Li \etal \cite{li2014Map} employed an ordered representation for hues, according to the sequence ordering of patches witnessed in clusters, with 0 for magenta, 1 for red, 2 for yellow, and so on. They defined the hue distances as the difference of these assigned values.

In order to test for a similar relationship between hue distances and the pattern of activities of model V4 neurons, we stacked our model V4 maps, in the order shown in Figure \ref{fig:network}, beginning with magenta, red, and so on. Stacking these maps results in a three-dimensional array, each column of which can be interpreted as a cluster of hue-selective patches, with neighboring patches sensitive to related hues, similar to those observed in V4 of monkeys \cite{li2014Map}. We call each column of our stacked maps a model cluster and each element of the column a model patch. An example of a model V4 cluster in a larger view is shown in Figure \ref{fig:network}. For a given model cluster and a pair of stimulus hues, we compute the distance of the two model patches within the cluster that are maximally activated by those hues. For example, the distance of the red patch from the cyan patch as shown in Figure \ref{fig:network} is 4. In this experiment, we employed our sampled hues from the HSL space, starting from red at 0 deg, separated 30 degrees, resulting in 12 stimulus hues. Similar to the ordering assigned to stimulus hues employed in \cite{li2014Map}, we assigned values in the range $[0, 5.5]$ at 0.5 steps starting with 0 for magenta. The plot in Figure~\ref{subFig:hue_distance} demonstrates our model patch distances as a function of hue distances with a clear correlation. The correlation coefficient was $r = 0.93, p = 2.09\times10^{-29}$. In other words, similar to the biological V4 cells, the pattern of responses in our model V4 neurons is highly correlated with the ordering of hues in the HSL space. 

\subsection*{Hue Reconstruction}
Li \etal~\cite{li2014Map} showed that in monkeys, 1-4 patches are needed to represent any hue in the visual field. Moreover, they showed that different hues were encoded with different multi-patch patterns. Then, they suggested that a combination of these activated patches can form a representation for the much larger space of physical colors. 

Along this line, we show, through a few examples, that for a given hue, a linear combination of model V4 neurons can be learned and used for representing that particular hue. It is important to note that it would be impossible to learn weights for the infinitely many possible physical hues. Hence, we show only a few examples here. However, our experiment is an instance of the possible mechanism for color representation suggested by Li~\etal~\cite{li2014Map}.

In this experiment, for a given hue value, we independently sampled the saturation and lightness dimensions at 500 points. The samples were uniformly distributed along each dimension. As a result, we have 500 colors of the same hue. The goal is to compute a linear combination of model V4 neurons, which can reconstruct the groundtruth hue. 

The hues in this experiment were represented as a number in the $(0, 2\pi]$ range. For numerical reasons, red is represented as $2\pi$, not 0. We performed an L1-regularized least square minimization, using the ``L1\_ls'' function described in \cite{Boyd_l1_ls}. 

Table~\ref{table:hue_exp_weights} shows some of the results for this experiment. Interestingly, in all cases, no more than four neuron types have large weights compared to the rest of the neurons. This is in agreement with the findings of~\cite{li2014Map}. Specifically, in the case of red and yellow hues, about 99\% of the contribution is from only a single cell, red and yellow neurons respectively. The last row in Table~\ref{table:hue_exp_weights} is most insightful. It presents the weights for a lavender hue in equal distance from blue (240 deg) and magenta (300 deg). The weights for this example seem counter-intuitive as they include green, cyan and magenta with positive contributions. In addition, blue is absent. However counterintuitive the weights seem, careful scrutiny of mean peak angles for V4 hues reveals that lavender hue at 270 deg is somewhere between the peaks for V4 cyan (at 193 deg) and magenta (at 300 deg), and closer to magenta. This hue is mainly reconstructed from magenta, with more than 70\% contribution, while the small weight for cyan is compensated with that of green. In other words, in this case, the green cell plays the role of shifting the reconstruction from magenta toward lavender.
\begin{table*}[t]
	\small\sf
	\centering
	\captionsetup{justification=centering}
	\caption{The choice of weights for model V4 cells used for hue reconstruction in a few example hues.\label{table:comb_experiment}}
	\begin{tabular}{ccccccc}
		\hline
		Groundtruth hue (deg) &\multicolumn{6}{c}{Model V4 neuron} \\
		\cline{2-7}
		& Red & Yellow & Green & Cyan & Blue & Magenta \\
		\hline
		red (360) &0.9903&0.0034 &0.0006&0.0007&0.0006&0.0014\\
		\hline
		Yellow (60) &0.0010&0.9950&0.0009&0.0021&0.0004&0.0005\\
		\hline
		lavender (270) &0.0001&0.0016&0.1960&0.0881&0.000&0.7142\\
		\hline
	\end{tabular}\\[10pt]
	\label{table:hue_exp_weights}
\end{table*}

Once again, it must be stressed that this experiment was performed to examine the possibility of combinatorial representation mechanisms and a thorough investigation of this mechanism in the computational sense is left for future work. The examples shown here attest to the fact that intermediary hues encoded by model V4 neurons can indeed span the massive space of physical hues and are enough for reconstructing any arbitrary hue from this space.

\section*{Discussion}
Our goal was to further understanding of the color processing mechanisms in the brain and to begin to assign color representational roles to specific brain areas. We investigated the contributions of each visual area LGN, V1, V2, and V4 in local hue representation by proposing a mechanistic computational model inspired by neural mechanisms in the visual system. Through a gradual increase in nonlinearity in terms of cone inputs, we observed a steady decrease in tuning bandwidths with a gradual shift in peak selectivities toward intermediate hue directions. Although one might be able to model the end result with a mathematical model in a single-layer fashion, such models do not lend insight to the neuronal mechanisms of color processing in the brain. In contrast, not only do our model neurons in each individual layer exhibit behavior similar to those of biological cells, but also at the system level, our hierarchical model as a whole provides a plausible process for the progression of local hue representation in the brain. The main difference in terms of potential insight provided by a single-layer mathematical model and our work is that our model can make predictions about real neurons that can be tested. A model whose contributions are of the input-output behavior kind cannot (see also \cite{Brown2014tale}).

We proposed multiplicative modulations in V2 as a means to increase nonlinearity in the hierarchy. We demonstrated that such modulations could rotate the cone-opponent axes to intermediate directions of perceptual red-green and yellow-blue hues and shift the tuning peaks toward unique hue angles. In short, our model predicts that multiplicative modulations are key operations in the encoding of hues in intermediate directions and unique hue representation. 

Our experimental results demonstrated that hue selectivity for model V4 neurons similar to that of neurons in area V4 of the monkey visual system was successfully achieved. Besides, our observations from the hue reconstruction experiment clearly confirmed the possibility of reconstructing the whole hue space using a combination of the hue-selective neurons in the model V4 layer. How this is achieved in the brain, for the infinitely many possible hues, remains to be investigated. 

Finally, our hierarchical network of neurons provides an important implication with regards to unique hue representations. Specifically, our computational experiments showed that as the visual signal moves through the hierarchy, responses with peaks close to unique hues start to develop. For unique red, our model single-opponent LGN cells peaked at less than a degree distance from this hue, while for unique green more complicated computations, or higher order mechanisms as put by others, were required and reaching such a close peak was delayed until model layer V2. Putting these together, we believe the answer to the question ``which region in the brain represents unique hues?'' is not limited to a single brain area, which in turn could be the source of disagreement among color vision researchers. Instead, our findings suggest that this question must be asked for each individual unique hue and that the answer will consist of an assorted set of brain regions for all four unique hues.

In our model, adding a variety of neurons such as concentric and elongated double-opponent color cells would result in a more inclusive system. However, we did not intend to make predictions about all aspects of color processing but only hue encoding mechanisms. We found that concentric double-opponent color cells, for example, have tuning bandwidth distribution similar to single-opponent neurons and tuning peaks along cone-opponent axes. This finding suggests that the contributions of concentric double-opponent cells are comparable to those of single-opponent neurons for hue representation, but we did not investigate those contributions in other color representations.

Our hierarchical model can be further extended to encode saturation and lightness. In the future, we would like to also address the problem of learning weights from V2 to V4. Furthermore, the experiment on hue reconstruction was performed with a simple linear regression model. A more sophisticated learning algorithm might result in more insightful weights. Lastly, in order to keep our model simple and avoid second-order equations, we skipped lateral connections between neuron types. However, these are part of the future development of a second-order model for our network.

\bibliographystyle{IEEEtran}
\bibliography{color_references}

\section*{Acknowledgements}
This research was supported by several sources for which the authors are grateful: Air Force Office of Scientific Research (FA9550-18-1-0054), the Canada Research Chairs Program, and the Natural Sciences and Engineering Research Council of Canada. The authors thank Dr. Oscar J. Avella and Dr. Rakesh Sengupta for useful discussions during the course of this research.

\section*{Methods}
In this work, the input to our model is LMS channels. In the event that the presented stimulus was available in RGB, we first performed a conversion into LMS channels using the transformation algorithm proposed by ~\cite{LMS_conversion} (we used the C code provided by the authors). As a result, one can think of the presented stimulus to the network as the activations of three cone types. These cone activations are fed to single-opponent LGN cells, which in turn feed single-opponent V1 cells with nonlinear rectification. In the V2 layer, single-opponent neurons replicate the activations of those of V1, but with larger receptive fields. Later, we refer to these single-opponent cells as ``additive V2 neurons''. ``Multiplicative V2'' neurons form when single-opponent V1 cells with L and M cone inputs are multiplicatively modulated by V1 neurons with S-cone input. This approach is in a sense similar to S-modulations proposed by De Valois \etal \cite{DeValois1993multi}, but in a multiplicative manner and not additive or subtractive. Finally, the hue-sensitive neurons in V4 receive feedforward signal from additive and multiplicative V2 cells.

Our model was implemented in TarzaNN~\cite{Rothenstein2005}. The neurons in all layers are linearly rectified. The rectification was performed using:
\begin{equation}
\phi(P) = 
	\begin{cases}
		\tau, & \text{if}\ mP+b < \tau\\
		mP+b , & \text{if}\ \tau \leq mP+b \leq s\\
		1, & \text{otherwise}, 
	\end{cases}
	\label{eq:rectifier}
\end{equation}
where $P$ is neuron activity, and $m$ and $b$ are the slope and base spike rate respectively, $\tau$ is a lower threshold of activities and $s$ represents the saturation threshold. This rectifier maps responses to $[\tau, 1]$. Depending on the settings of parameters $\tau$ and $s$, and the range of activations for the model neurons, the rectifier might vary from being linear to nonlinear. Wherever this rectifier is employed in the rest of the paper, we mention the settings of the parameters, and whether parameter settings resulted in neuron activations to become linear or nonlinear in terms of their input.

The input to the hierarchical network was always resized to $256\times256$ pixels. The receptive field sizes, following~\protect\cite{freeman2011metamers}, double from one layer to the one above. Specifically, the receptive field sizes we employed were $19\times19$, $38\times38$, $76\times76$, and $152\times152$ pixels for LGN, V1, V2, and V4 layers respectively.

\subsection*{Model LGN Cells}
The first layer of the hierarchy models single-opponent LGN cells. The LGN cells are characterized by their opponent inputs from cones. For example, LGN cells receiving excitatory input from L cones and inhibitory signals from M cones are known as L-on cells. Model LGN cell responses were computed by~\cite{shapley2011color}:
\begin{eqnarray}
R_\text{LGN} & = \phi(& a_\text{L} (G(x, y, \sigma_\text{L}) \ast R_\text{L}) +\nonumber\\
			    &    & a_\text{M} (G(x, y, \sigma_\text{M}) \ast R_\text{M}) +\nonumber\\
			    &    & a_\text{S} (G(x, y, \sigma_\text{S}) \ast R_\text{S})), 
\label{eq:LGN}
\end{eqnarray}
where $\ast$ represents convolution. In this equation, model LGN response, $R_\text{LGN}$, is computed 
by first, linearly combining cone activities, $R_\text{L}$, $R_\text{M}$, and $R_\text{S}$, convolved with normalized Gaussian kernels, $G$, of different standard deviations, $\sigma$, followed by a linear rectification, $\phi$. For model LGN neurons, we set $\tau = -1$ and $s=1$ to ensure the responses of these neurons are linear combinations of the cone responses~\cite{DeValois2000physio,Lennie1990chromatic}. The differences in standard deviations of the Gaussian kernels ensure different spatial extents for each cone as described in~\cite{Reid2002}. Each weight in Eq.~\ref{eq:LGN}, determines presence/absence and excitatory/inhibitory effect of the corresponding cone. The weights used for model LGN cells were set following~\cite{Reid2002} and~\cite{johnson2004cone}. In total, we modeled six different LGN neuron types, L-on, L-off, M-on, M-off, S-on, and S-off. As an example, consider the weights for M-on cells as $-1.0, 1.1, 0$ from L, M, and S cones respectively. These neurons receive opposite contributions from L and M cones, while S cones with weight 0 exhibit no contribution. That is, M and L cones have excitatory and inhibitory effects respectively, while S cones are absent. This type of neuron is known to best respond to cyan-like hues \cite{Conway2001spatial}. A relatively similar hue selectivity is observed in L-off cells, with $-1.1, 1.0, 0$ weights from L, M, and S cones respectively. L-off cells, similar to M-on single-opponent neurons, receive excitatory and inhibitory contributions from M and L cones respectively. Receiving such a pattern of input signal results in selectivities to cyan-like hues for both cell types. Despite this similarity, however, their responses indicate different messages. Specifically, strong responses of an M-on cell conveys high confidence in the existence of an M cone signal with less impact from L cone responses within its receptive field. In contrast, strong activations of an L-off cell indicates the existence of M cone signal with a confident message that almost no L cone activities exist within its receptive field. In other words, even a small amount of L cone responses within the receptive field of an L-off cell strongly suppresses the activation of this neuron. Looking at the feature maps for these two cell types in Figure \ref{fig:qual_examples} reveals these slight differences in their selectivities. Note that while both neurons have relatively strong positive responses to the green and blue regions in the input, activations of the L-off cells, unlike the responses of M-on neurons, in the yellow region are strongly suppressed.
 
In what follows, whenever we refer to a cell as L, M, or S in layers LGN and higher, we will be referring to the pair of on and off neurons in that layer. For instance, M-on and M-off neurons in LGN might be called M neurons in this layer, for brevity.
 
\subsection*{Model V1 cells}
Local hue in V1, as suggested in~\cite{shapley2002neural} and~\cite{johnson2004cone}, can be encoded by single-opponent cells. To obtain such a representation in the model V1 layer, the responses are determined by convolving input signals with a Gaussian kernel. Note that since single-opponency is implemented in the model LGN layer, by simply convolving model LGN signals with a Gaussian kernel, we will also have single-opponency in  V1. The local hue responses of V1 were obtained by:
\begin{equation}
R_{\text{V1}} = \phi(G(x, y, \sigma_{\text{V1}}) \ast R_\text{LGN}),
\label{eq:V1_resp}
\end{equation}
where $\phi$ is the rectifier in Eq.~\ref{eq:rectifier}. With $\tau = 0$ and $s = 1$ for the rectifier, our model V1 neurons will be nonlinear functions of cone activations. In Eq.~\ref{eq:V1_resp}, substituting $R_\text{LGN}$ with any of the six model LGN neuron type responses will result in a corresponding V1 neuron type. Therefore, there are six neuron types in layer V1 corresponding to L-on, L-off, M-on, M-off, S-on, and S-off.

The size of the Gaussian kernels for each of these neurons determines their receptive field sizes. In our implementation, the receptive field size doubles from one layer to the next following similar observations in the ventral stream~\cite{freeman2011metamers}.

\subsection*{Model V2 cells}
In our network, the V2 layer consists of two types of hue selective cells: single-opponent and multiplicative. The single-opponent neurons are obtained by:
\begin{equation}
R_{\text{additive V2}} = \phi(G(x, y, \sigma_{\text{V2}}) \ast R_{\text{V1}}),
\label{eq:V2_resp}
\end{equation}
where $\phi$ is the rectifier in Eq.~\ref{eq:rectifier}. With $\tau = 0$ and $s = 1$ for the rectifier, the single-opponent V2 cells are nonlinear functions of cone activations. In Eq.~\ref{eq:V2_resp}, substituting $R_{\text{V1}}$ with each of the six model V1 neuron type responses will yield a model V2 neuron type with similar selectivities, but with a larger receptive field. To be more specific, the responses of single-opponent V2 neurons can be considered as a linear combination of V1 activations.

To increase the nonlinearity as a function of cone activations in V2, as observed by Hanazawa \etal \cite{Hanazawa}, and also to nudge the selectivities further toward intermediate hues, as found by Kuriki \etal \cite{Kuriki2015fMRIhue}, we introduce multiplicative V2 neurons. These cells not only add another form of nonlinearity to the model, other than that obtained by the rectifier in V1, but also mix the different color channels from V1 and exhibit a decrease in their tuning bandwidths. In their model, De Valois \etal \cite{DeValois1993multi} suggested that S-modulated neurons rotate the cone-opponent axes to perceptual-opponent directions. Their modulations with S activations were in the form of additions and subtractions, which does not add to the nonlinearity of neuron responses. We leverage their observation, but in the form of multiplicative modulations for additional nonlinearity. That is, each V2 multiplicative cell response is the result of multiplying L or M neurons from V1 with a V1 S cell activations. For example, in Figure \ref{fig:qual_examples}, ``L-off $\times$ S-off'' is for a cell obtained by modulating a V1 L-off cell responses by a V1 S-off neuron activations.  In our model, the multiplicative V2 neurons are computed as:
\begin{eqnarray}
R_{\text{multiplicative V2}} & = & \phi(G(x, y, \sigma_{\text{V2}}) \ast R_{\text{V1\{L, M\}}} \times \nonumber\\
&& G(x, y, \sigma_{\text{V2}}) \ast R_{\text{V1\{S\}}}),
\label{eq:V2_multi_resp}
\end{eqnarray}
where $\times$ represent multiplicative modulation, and $R_{\text{V1\{L, M\}}}$ and $R_{\text{V1\{S\}}}$ are for responses of an L or M cell and S neuron from V1 respectively. As before, $\phi$ is the rectifier from Eq. \ref{eq:rectifier} with the same parameters as those of the additive V2 cells. Multiplicative V2 cells are nonlinear with respect to cone inputs and bilinear with regards to V1 activations.

Multiplicative V2 neurons have narrower bandwidths than those of additive V2 cells, which we showed quantitatively earlier. However, consider the multiplicative V2 maps in Figure \ref{fig:qual_examples} for a brief qualitative explanation. For the hue wheel as input, relatively high responses of the single-opponent V2 cells span a larger region of their map compared to multiplicative V2 cells. This is an indication that multiplicative V2 cells are selective to a narrower range of hue angles. As an example, both L-off and S-off V1 cells have high activations for relatively large regions of the input respectively. However, when multiplied, the resulting neuron, \ie the ``L-off $\times$ S-off'' cell has strong responses for regions with greenish color, and the activation of the L-off V1 cell to bluish regions is suppressed to the extent that the L-off $\times$ S-off cell shows close to no responses.

As a summary, in model layer V2, a total of 14 neuron types are implemented: 6 additive and 8 multiplicative cell types.

\subsection*{Model V4 cells.}
We modeled V4 neurons representing local hue using a weighted sum of convolutions over model V2 neuron outputs. More specifically, responses of the $i$-th V4 neuron, $R_{\text{V4}, i}$, are computed as:
\begin{equation}
R_{\text{V4}, i} = \phi(\sum_{j=1}^{14} w_{ij}(G(x, y, \sigma_{\text{V4}}) \ast R_{\text{V2}, j})),
\label{eq:V4}
\end{equation}
where $R_{\text{V2}, j}$ represents the responses of the $j$-th V2 neuron, and $\phi$ is the rectifier introduced in Eq.~\ref{eq:rectifier}, with $\tau = 0$ and $s = 1$. As a result of this parameter setting for the rectifier, each V4 cell is a linear combination of V2 cell responses and hence, nonlinear in terms of cone inputs. The set of weights $\left\{w_{ij}\right\}_{j = 1, \ldots, 14}$ determine the hue to which the $i$-th model V4 neuron shows selectivity.

In model layer V4, we implemented six different neuron types according to distinct hues: red, yellow, green, cyan, blue, and magenta. The chosen hues are 60 deg apart on the hue circle of HSL, with red at 0 deg. These hues were also employed in the V4 color map study~\cite{li2014Map} and for comparison purposes, we utilize these hues. When the six V4 colors are mapped to the MB space, the hue angles are shifted with respect to those of HSL, with red, yellow and cyan hues close to cone-opponent directions in the MB space, and green, blue and magenta along off-opponent axes. From here on, we will refer to V4 neurons based upon their selectivities, e.g., model V4 red or model V4 cyan neurons. Although here we limit the number of modeled neuron types in this layer to six, we would like to emphasize that changes in combination weights will lead to neurons with various hue selectivities in this layer. Modeling neurons with selectivities to a wide variety of hues with yet narrower tunings could be accomplished in higher layers, such as IT, by combining hue-selective model neurons in V4. 

In order to determine the weights from V2 to V4 neurons, $w_{ij}$'s in Equation \ref{eq:V4}, we considered the distance between mean peak activations of model V2 neurons to the desired hue in a model V4 cell. The hue angle between these two hues on the hue circle is represented by $d_{ij}$. Then, the weight $w_{ij}$ from model V2 neuron $j$ to model V4 neuron $i$ is determined by:
\begin{equation}
w_{ij} = \frac{\mathcal{N}(d_{ij}; 0, \sigma)}{Z_i},
\label{eq:V4_weights}
\end{equation}
where $\mathcal{N}(.;0, \sigma)$ represents a normal distribution with 0 mean and $\sigma$ standard deviation, and $Z_i$ is a normalizing constant obtained by
\begin{equation}
Z_i = \sum_{j=1}^{14}\mathcal{N}(d_{ij}; 0, \sigma).
\end{equation}
The weights used for each of V4 neuron types are summarized in Figure \ref{subfig:V2_V4_weights}.
In this figure, each row represents the weights for a single V4 cell, and the columns are for model V2 cells. Note that all V2 to V4 weights are normalized to sum to 1.0. That is, the sum of weights in each row is 1. In this figure, dark red shows a large contribution, while dark blue represents close to no input from the relevant V2 neuron. Consider, for example, the weights for the red V4 cells. This neuron has relatively large weights from V2 L-on, M-off, and M-off $\times$ S-off cells. In other words, cells with large contributions from L cones. This observation is not surprising as previous research by Webster~\etal~\cite{Webster2000variations} found that unique red in human subjects has largest contributions from L cones. 

In Figure~\ref{fig:qual_examples}, at the V4 layer, from top to bottom, the neurons selective to magenta, red, yellow, green, cyan, and blue are displayed. As expected, model V4 yellow neurons, for instance, show activations across red, yellow, and green regions of the stimulus, with stronger activations in the yellow segment.

\subsection*{Choice of the model}
Looking back at our network architecture in Figure~\ref{fig:network}, and also the computational operations for each layer of our model, the reader might wonder why we did not employ a convolutional neural network (CNN) for hue representation. After all, our network architecture is similar to that of a CNN: the responses of neurons in each layer of the model are computed by a convolution followed by a rectification, similar to the operations in a CNN. We emphasize here that our choice of the model differs from a CNN for the following reasons:
\begin{enumerate}
	\item Our goal was to introduce a biologically inspired model that would help in understanding hue encoding mechanisms in the brain. In doing so, we designed each neuron in our network according to the existing findings of the brain. For example, the receptive field profile and the weights from cones to single-opponent cells in our model LGN layer were set based on the reported findings of Reid \etal \cite{Reid2002}. In a CNN, these parameters of the model are learned from data, and as a result, any receptive field profile and any setting of weights might be learned, which could possibly be different from those of biological color neurons. Similar to our discussion about one-layer models, in an end-to-end manner, CNNs might succeed in hue representation and specifically in encoding of unique hues. However, the individual neurons in such models might not match with those of the brain and hence, will not demystify color processing in the brain.
	\item One challenge in convolutional neural networks is interpreting the learned features in the hidden layers. Often, the learned features in the first hidden layer are compared with biological V1 neurons. However, learned features in deeper layers are difficult to explain. There have been attempts to understand and interpret hidden layer features \cite{networkDissection, olah2018building}. However, a clear understanding of learned features and the ability to explain the reason behind the decision in learning those features is yet to be achieved. As a result, had we employed a CNN model, we would not have been able to explain the learned features in all layers of our model, which was far from our goal to assign representational roles to each brain area from LGN to V4.
	\item In this work, we did not have access to any cell recording data. Nonetheless, even with such data accessible to us, we would not have been able to use a CNN model. Often, cell recording data is limited and sparse and not enough for learning the massive number of parameters in a CNN.  
	\end{enumerate}
	We acknowledge that a certain set of parameters in our model were set according to biological findings and the remaining parameters, such as the weights from our model layer V2 to V4, were set heuristically. Indeed, a learning algorithm, in this case, might prove to help make predictions about these connections in the brain. This step, as described in the Discussion section, is left to be further explored in the future.
\end{document}